\documentclass{article}
\usepackage{amsbsy} 
\usepackage{booktabs}	
	\usepackage{multirow} 
	\usepackage{multicol}
\usepackage{comment}  
\usepackage{color}
\usepackage{multicol}
\usepackage{graphicx}
\graphicspath{{./}} 
\usepackage[version=3]{mhchem} 
\usepackage{setspace}
\usepackage{subfigure}
\usepackage{url}
\usepackage{xspace}
\usepackage{lastpage} 
\usepackage[left=3cm,bottom=3cm,top=3cm,right=3cm]{geometry}

\usepackage{fancyhdr}
\fancypagestyle{plain}{ %
\fancyhf{} 
\setlength{\headheight}{15.2pt}
\setlength{\footskip}{40pt}
\lhead{\textit{Silver segregation to \thetaprime-Al interfaces in Al-Cu-Ag alloys}}
\rhead{J.M.Rosalie\ \& L. Bourgeois}
\lfoot{\textit{Pre-print}}
\rfoot{\thepage\ of\ \pageref{LastPage}}

}

\usepackage[plainpages=false,
pdftitle={Silver segregation to thetaprime (Al2Cu)-Al interfaces in Al-Cu-Ag alloys},
pdfauthor={Julian M. Rosalie,  Laure Bourgeois}
pdfsubject={Aluminium alloys Interfaces, Nucleation,Segregation},
pdfkeywords={Aluminium alloys, Precipitation, Interfaces,Segregation},
colorlinks,
debug,
linkcolor=red,
citecolor=red, 
filecolor=blue,
breaklinks=true]{hyperref}

\newcommand{\gammaprime}{\ensuremath{\gamma^{\prime}}\xspace}
\newcommand{\thetaprime}{\ensuremath{\theta^{\prime}}\xspace}

\begin{document}

\title{Silver segregation to \thetaprime (\ce{Al2Cu})-Al interfaces in Al-Cu-Ag alloys} 
\author{Julian~M.~Rosalie$^{a}$ 
\and Laure~Bourgeois$^{b,c,d}$}
\date{}

\maketitle
\noindent
$^a$Structural Materials Unit, National Institute for Materials Science, Tsukuba, 305-0047, Japan.\\
$^b$Monash Centre for Electron Microscopy, Monash University, 3800, Victoria, Australia.\\
$^c$ARC Centre of Excellence for Design in Light Metals, Australia.\\
$^d$Department of Materials Engineering, Monash University, 3800, Victoria, Australia.\\

\begin{abstract} 

\thetaprime (\ce{Al2Cu}) precipitates in Al-Cu-Ag alloys were examined using high angle annular dark field  scanning transmission electron microscopy (HAADF-STEM). 
The precipitates nucleated on dislocation loops on which assemblies of \gammaprime (\ce{AlAg2}) precipitates were present. 
These dislocation loops were enriched in silver prior to \thetaprime precipitation.
Coherent, planar interfaces between the aluminium matrix and \thetaprime precipitates were decorated by a layer of silver of two atomic layers in thickness.
It is proposed that this layer lowers the chemical component of the Al-\thetaprime interfacial energy.
The lateral growth of the \thetaprime precipitates was accompanied by the extension of this silver bi-layer, resulting in  the loss of silver from neighbouring \gammaprime precipitates and contributing to the deterioration of the \gammaprime precipitate assemblies. 

\end{abstract}



\section{Introduction}

Aluminium-copper (Al-Cu) alloys are one of the most-studied precipitation strengthened alloy systems, in part due to their historic connection to the first precipitation-hardened Al alloys (Al-Cu-Mg) \cite{Wilm:1911, GayleAluminium1994}.
Although Al-Cu alloys are amongst the earliest examples of precipitation-strengthened Al systems, alloys derived from these  compositions, such as alloy designations 2014  and 2024, are still used in aerospace applications \cite{ZhuAluminium2002}.

Al-Cu alloys derive a significant measure of their strength from the controlled precipitation of the \thetaprime phase on the \{100\} planes of the Al matrix.  
These precipitates impede dislocation glide, enhancing the resistance to plastic deformation and hence increasing the yield strength.
The Al-Cu system is also of great practical interest since the precipitation of the \thetaprime phase can be manipulated through deformation and by the presence of various solutes, some of which (such as In, Sn and Cd) are potent even at trace levels \cite{BourgeoisSn2011,silcock:1955,silcock:1955a}. 

Industrial variants of these alloys have considerably more complex compositions than simple, ``model'' alloys. 
Compositions are frequently chosen so as to improve mechanical properties through the formation of multiple precipitate phases. 
In Al-Cu-Mg alloys, for example, laths of S phase (\ce{Al2CuMg}) can form in addition to \thetaprime plates. 
Multiple phase precipitation and the presence of additional elements in solution pose severe difficulties in clarifying the effect of any given  element on \thetaprime precipitation. 
Consequently, despite research spanning the past half-century, the mechanism(s) by which solute elements affect the precipitation of the \thetaprime phase remain poorly understood. 

The precipitate-matrix interface plays an important role in precipitation strengthening. 
A well-known example of this involves additions of Ag to Al-Cu-Mg alloys, which results in the precipitation of a phase termed $\Omega$ rather than \thetaprime.
Although the $\Omega$ phase shares the same bulk composition as \thetaprime, it forms with \{111\} habit instead of \{100\} and displays remarkable coarsening resistance. 
Drift corrected energy dispersive x-ray analysis \cite{OkunishiOmega2001} and 
3D atom probe studies \cite{HonoOmega1993} have shown that Ag and Mg both segregate to the $\Omega$-Al matrix interface.   
The $\Omega$ phase has a distinctive interface structure, with a bi-layer of Ag atoms at the  coherent  interface with the matrix\cite{HutchinsonOmega2001}.
Density functional theory calculations indicated while substitution of either Mg or Ag alone was energetically unfavourable, a combination of both elements provided a lower-energy interface \cite{SunOmega2009}. 

Recent work has shown that compositional changes at the \thetaprime-matrix interface occur in other alloys.  
Silicon was found to segregate to the coherent \{100\} interfaces of the \thetaprime phase in Al-Cu-Si alloys, substituting for Cu sites in the precipitate \cite{BiswasThetaprime2011}.
It has also been demonstrated that the interface composition and structure of \thetaprime in binary Al-Cu differs substantially from the bulk structure, with additional Cu atoms occupying octahedral interstices in the precipitate \cite{BourgeoisAlCu2012}. 
Despite these studies a detailed understanding of precipitate-matrix interfaces has yet to be developed in the majority of Al alloys.

Aluminium-copper-silver (Al-Cu-Ag) alloys are ideal for studying the effect of a third element on the precipitation of the \thetaprime phase.   
Unlike many ternary systems (e.g. Al-Cu-Mg) the Al-rich region of the phase field contains only binary phases \cite{liu:1983}, eliminating the complexities associated with competing ternary phase precipitation. 
The interatomic spacing in pure Ag also corresponds very closely to that of pure Al, and substantial additions of silver have little effect on the lattice parameter of Al \cite{hume-rothery1940,vegard:1921}. 
Solute misfit can therefore be regarded as minimal, and solute-induced strain can be largely neglected.
In addition, the chemical affinity between Ag and Cu atoms is weak and the strong co-clustering behaviour exhibited by systems such as Al-Si-Mg is not observed.

Early studies on Al-Cu-Ag alloys established that for compositions of $\sim$1at.\%(Ag,Cu)  both \gammaprime (\ce{AlAg2}) and \thetaprime precipitates were formed \cite{bouvy:1965} and it been shown that both phases can form on dislocation loops, with \thetaprime and \gammaprime platelets observed at different regions of the same dislocation loop \cite{Rosalie2009b,RosalieTms2012Thetaprime}.

We have recently shown that there is significant silver segregation around the dislocation loops at the site where \thetaprime precipitates nucleate and that the precipitates are later surrounded by an atmosphere of silver \cite{RosalieTms2012Thetaprime}. 
This work examines the segregation of Ag to \thetaprime precipitates in a Al-Cu-Ag alloy with an emphasis on the atomic structure of the Al-\thetaprime coherent interfaces.

\section{Experimental details}

Alloys containing Al-0.90at.\%Ag-0.90at.\%Cu	(Al-3.45wt.\%Ag-2.05wt.\%Cu) were cast from high-purity elements in air at 700$^\circ$C.
The composition was verified via inductively coupled plasma-mass spectroscopy (ICP-MS). 
The largest single impurity was Fe (0.02wt.\%), with most other impurities (including Mg) present at $<$0.005wt.\%.
The cast ingots were homogenised (525$^\circ$C, 168\,h) before being hot-rolled (to 2\,mm thickness), followed by cold-rolling to 0.5\,mm thickness.
Further external deformation was avoided from this point onwards.
Discs (3mm diameter) were punched from the sheet, solution-treated (525$^o$C, 0.5\,h) in a nitrate/nitrite salt pot and quenched to room temperature in water. 
Isothermal ageing of these discs was conducted in an oil bath at 200$^\circ$ for between 2 and 4\,h. 

TEM foils were prepared by manually grinding  the aged discs, followed by thinning to perforation by twin jet electropolishing in a nitric acid/methanol solution ($\sim$13\,V, $-20^\circ$C in  33\% \(\mathrm{HNO_3}\)$/$67\%\(\mathrm{CH_3OH}\) v/v).
Discs were plasma-cleaned prior to examination.

High angle annular dark field (HAADF) scanning transmission electron microscope (STEM) images were obtained using a 
FEI Titan$^3$ 80-300 microscope operating at 300\,kV,  with  convergence semi-angle of 15\,mrad thus resulting in a spatial resolution of $\approx$ 1.2\,\AA~and with an inner collection angle of 40\,mrad.  
By default, images are presented adjusted only for brightness and contrast.
Image analysis was performed using ImageJ software, version 1.44o.
EDX maps were obtained in STEM mode using a JEOL Si(Li) detector installed on a JEOL 2100F microscope operating at 200\,kV with a probe size of 0.5\,nm. 
The maps were 256x256 pixels in size, with an acquisition time of 0.2\,ms per pixels over $\approx$ 100 frames.

\section{Results}

\subsection{Microstructural overview}
In foils aged for 0.16$-$2\,h at 200$^\circ$C the microstructure contains only \gammaprime (\ce{AlAg2}) precipitates and   Ag-rich Guinier Preston zones (GPZ). 
The \gammaprime precipitates are arranged in elliptical loops, each comprised of alternating variants, with the loop normal close  to a $\langle 011 \rangle$ direction. 
The morphology and formation of these precipitate assemblies have been described previously \cite{Rosalie2009b} and are therefore not discussed in detail. 

\begin{figure*}
	\begin{center}	
				\hfill
				\subfigure[4h at 200$^\circ$C \label{fig-overview1}]{\includegraphics[width=5.5cm]{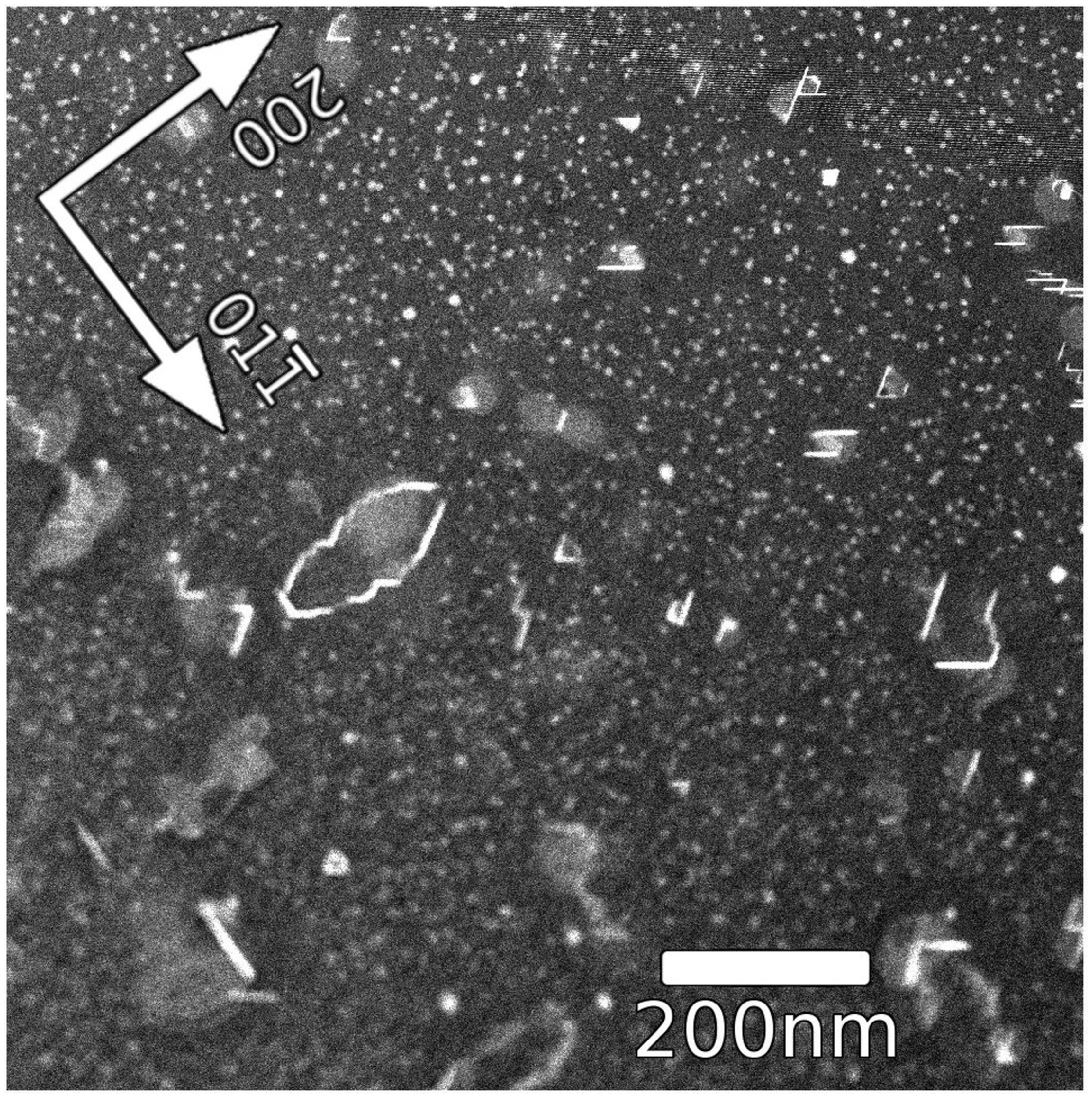}}
				\hfill
				\subfigure[4h at 200$^\circ$C \label{fig-overview2}]{\includegraphics[width=5.5cm]{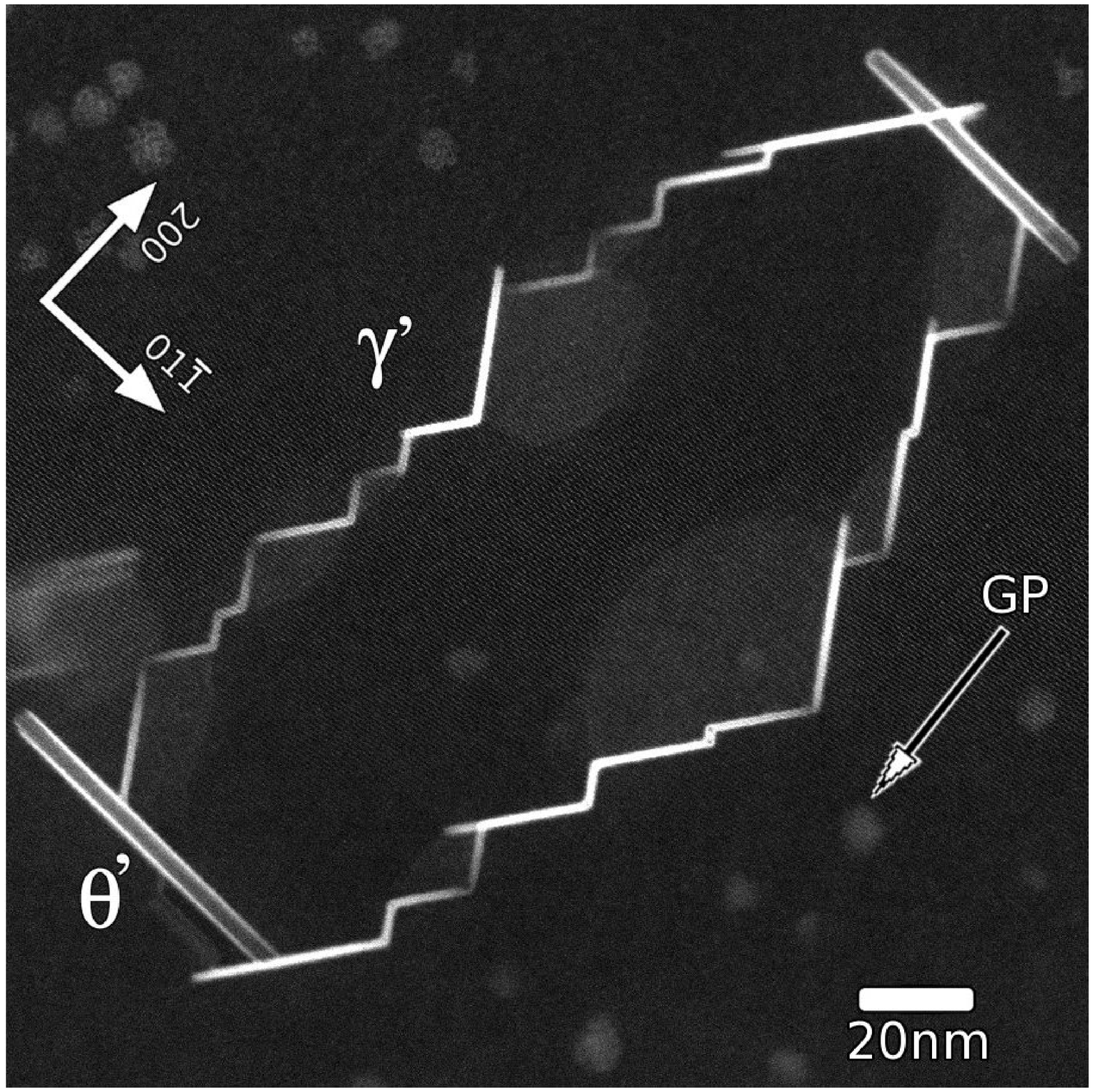}}
				\hfill\
		\caption{HAADF-STEM micrographs illustrating the microstructure of the Al-Cu-Ag alloy aged for 4\,h at 200$^\circ$C.
(a) Low magnification  image showing several precipitate assemblies (some truncated in foil preparation). 
(b) shows a single, similar assembly in greater detail with two \thetaprime and a large number of \gammaprime precipitates present.  
Spheroidal Ag GP zones can be seen around the assemblies in both images. 
 \label{fig-overview}}
	\end{center}
\end{figure*}

\thetaprime precipitates are observed in samples aged for a minimum of 2\,h at 200$^\circ$C. 
Figure~\ref{fig-overview1} presents a low magnification HAADF-STEM image obtained after 4\,h ageing and shows several  \gammaprime precipitate assemblies, some of which also include \thetaprime precipitates.
Spheroidal Ag-rich GPZ are also present. 
Most \gammaprime assemblies have been truncated during foil preparation; however, one complete example is present in the centre of the image. 

 \thetaprime precipitates are frequently observed in pairs at opposite ends of a precipitate assembly, as is shown in Fig~\ref{fig-overview2}.
Both \thetaprime precipitates in Fig~\ref{fig-overview2} show enhanced contrast on the coherent interfaces with the matrix, which, as will be demonstrated, is due to silver segregation. 

\begin{figure*}
	\begin{center}
		\hfill
		\subfigure[HAADF-STEM\label{edx-theta-Ag}]{\includegraphics[width=0.3\textwidth]{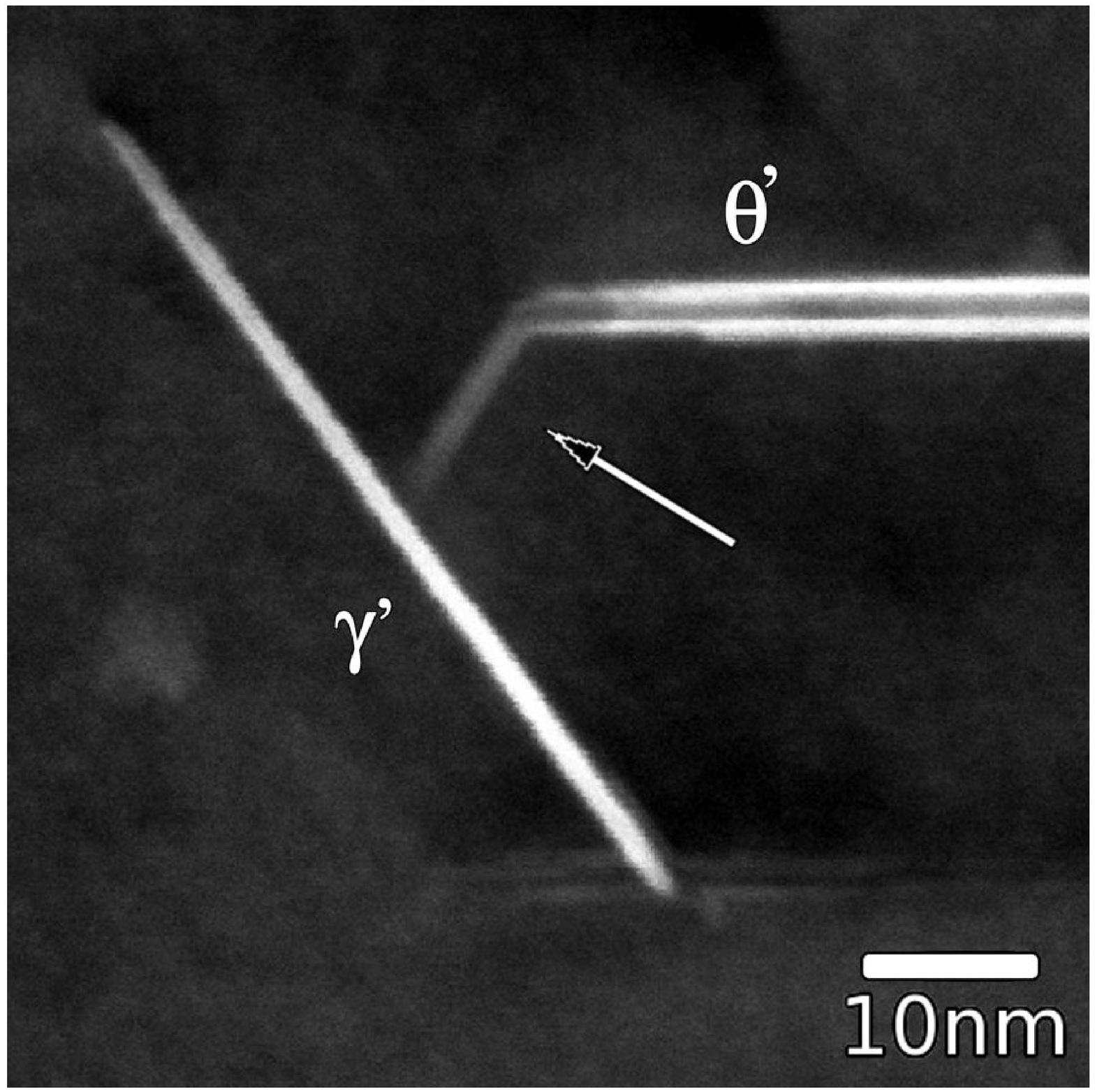}}
		\hfill
		\subfigure[Ag L \label{edx-theta-Ag}]{\includegraphics[width=0.3\textwidth]{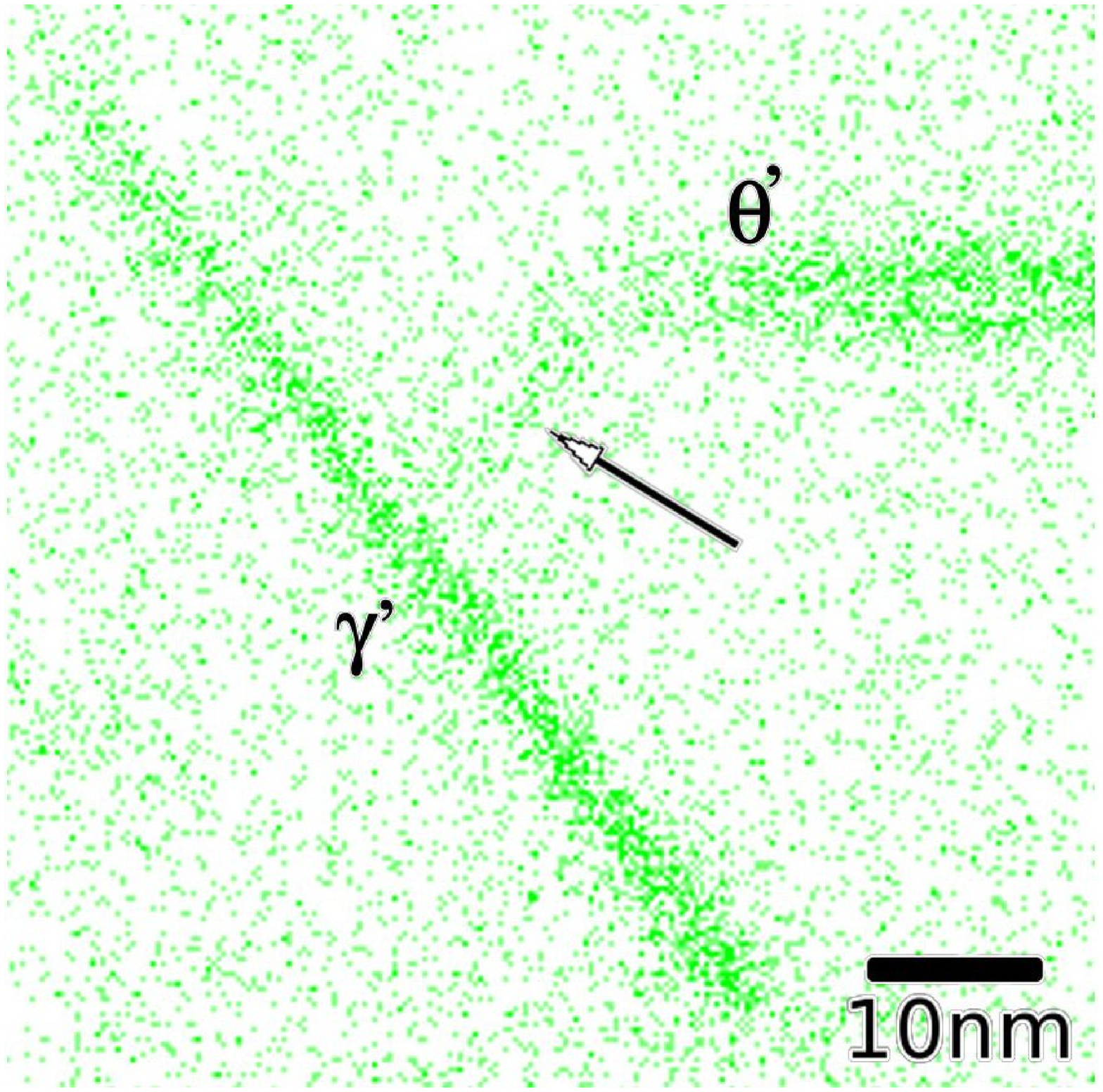}}
		\hfill
		\subfigure[Cu K\label{edx-theta-Cu}]{\includegraphics[width=0.3\textwidth]{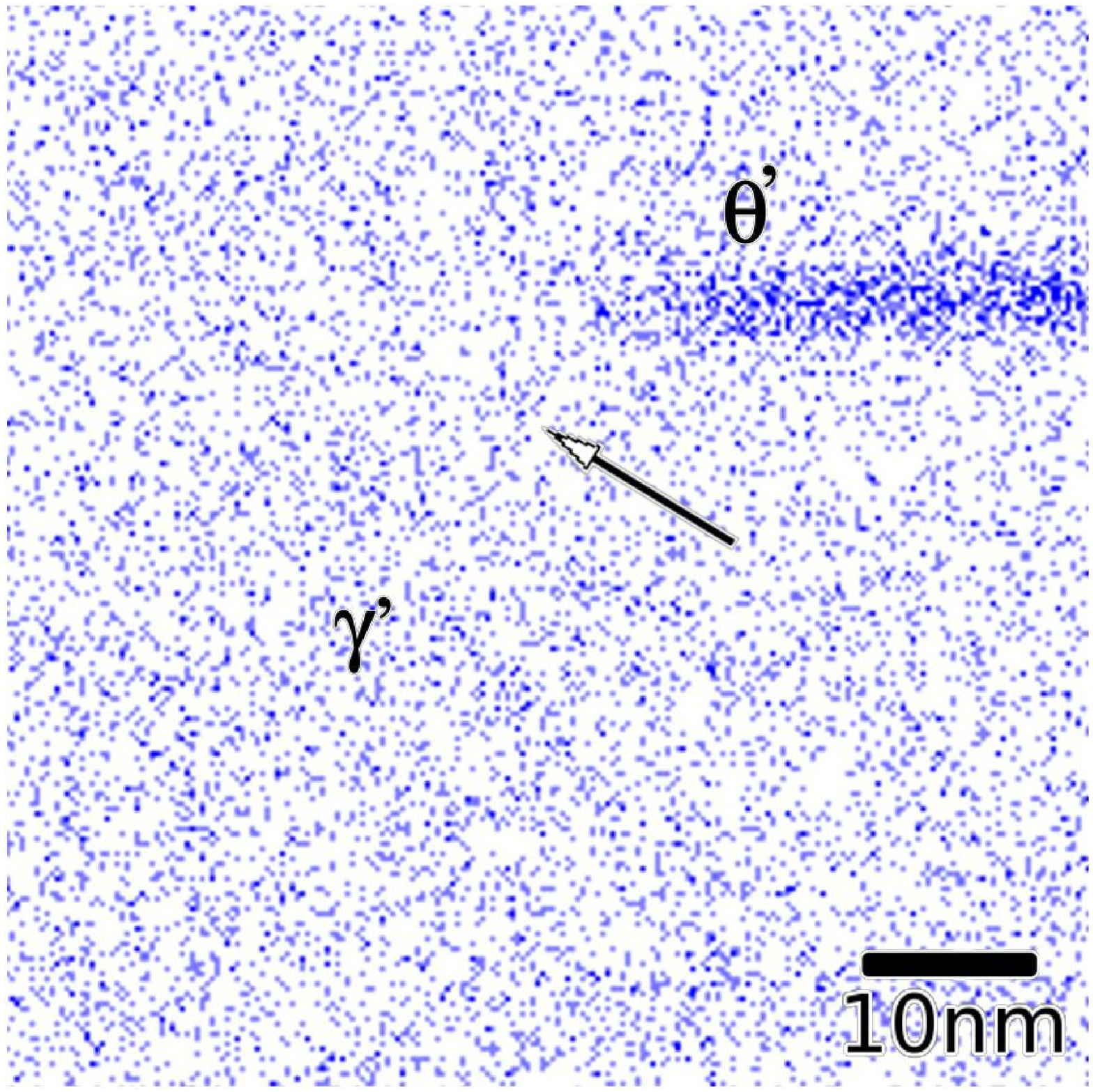}}
		\hfill\

		\caption{(a)HAADF-STEM image and (b,c) EDX maps of a  \gammaprime-\thetaprime precipitate assembly.  The Ag map (b) shows silver at the sites of both precipitates and along the defect  (see arrow) between them. 
Cu is concentrated only at the site of the \thetaprime precipitate. 
 \label{theta-edx}}
	\end{center}
\end{figure*}

EDX maps show segregation of silver to residual dislocations at the ends of the precipitate assemblies.
Figure~\ref{theta-edx} shows HAADF-STEM images and EDX maps of \thetaprime and \gammaprime precipitates at the end of an assembly. 
Both precipitates display strong contrast in the HAADF image, as expected from the higher atomic contrast of Cu and Ag compared to Al. 
Between both phases a diffuse line of contrast can be seen (indicated by an arrow).
EDX maps based on the Ag L  peak (Fig~\ref{edx-theta-Ag}) show that Ag is present at the sites of both \gammaprime and \thetaprime phases.
Ag is also concentrated along the lattice defect between \gammaprime precipitates. 
The EDX map based on the Cu K peak (Fig~\ref{edx-theta-Cu}) shows Cu only at the site of the \thetaprime precipitate and not at  \gammaprime or the lattice defect. 
Similar EDX observations of silver segregation to dislocation loops adjacent to \gammaprime precipitate assemblies have also been reported recently \cite{RosalieTms2012Thetaprime}.

\begin{figure*}
	\begin{center}
		\subfigure[HAADF-STEM image\label{defect-HAADF-STEM}]{\includegraphics[width=0.48\textwidth]{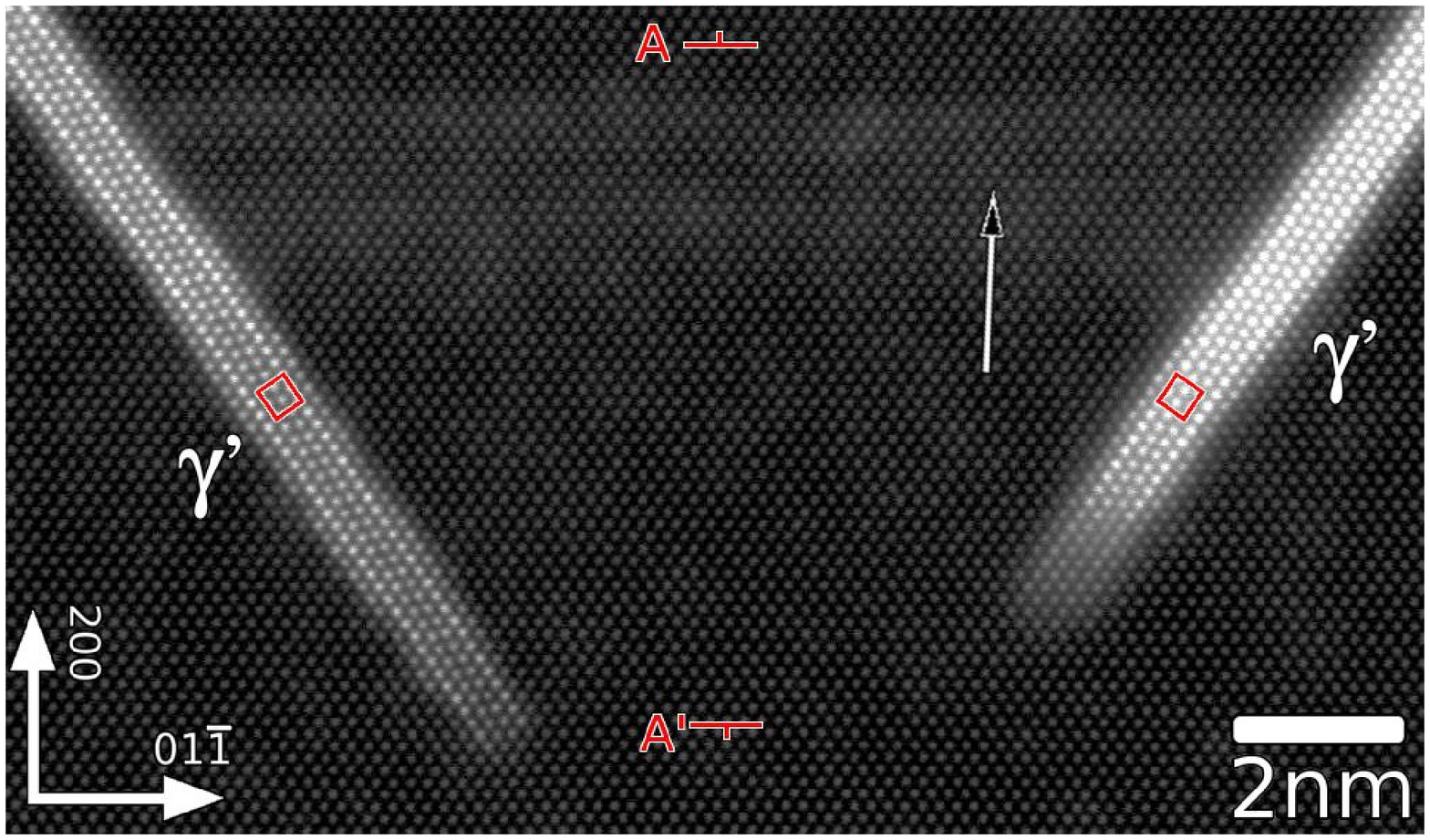}}\hfill
		\subfigure[HAADF-STEM intensity along A-A'\label{defect-line}]{\includegraphics[width=0.48\textwidth]{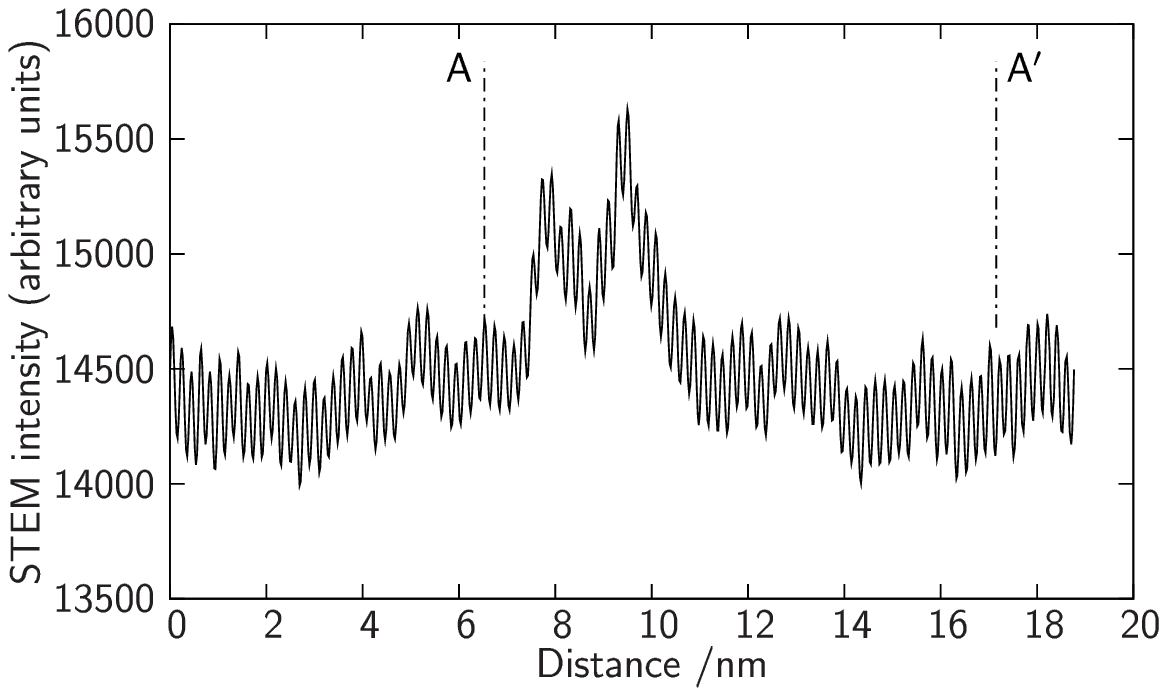}}
		\caption{\gammaprime precipitate assembly prior to \thetaprime phase nucleation.
HAADF-STEM images (a) and intensity profiles (b) show a diffuse band of higher atomic contrast elements running between the \gammaprime precipitates.
The defect (arrowed) is apparent in increased atomic contrast between the \gammaprime precipitates.
The HAADF-STEM intensity (d) indicates that the solute Ag is enriched in two bands.
\gammaprime unit cells are indicated by outlines on the micrograph. 
\label{edx-defect}}
	\end{center}
\end{figure*}

These silver-enriched residual defects are also present before \thetaprime phase precipitation. 
The HAADF-STEM image presented in Figure~\ref{defect-HAADF-STEM} shows two \gammaprime precipitates  with edges separated by approximately 10\,nm. 
There is a weak diffuse line (indicated by the arrow) between the \gammaprime precipitates indicating the presence of higher atomic number solutes.  
A plot of the HAADF-STEM intensity along $A$--$A'$ (Fig.~\ref{defect-line}) shows stronger atomic contrast in two bands, separated by approximately 1.5\,nm.

\subsection{\thetaprime-Al interfaces}

\subsubsection{Coherent \thetaprime-Al interfaces}
\begin{figure*}
	\begin{center}
		\hfill
		\subfigure[HAADF-STEM\label{fig-theta-stem-1}]{\includegraphics[width=5.5cm]{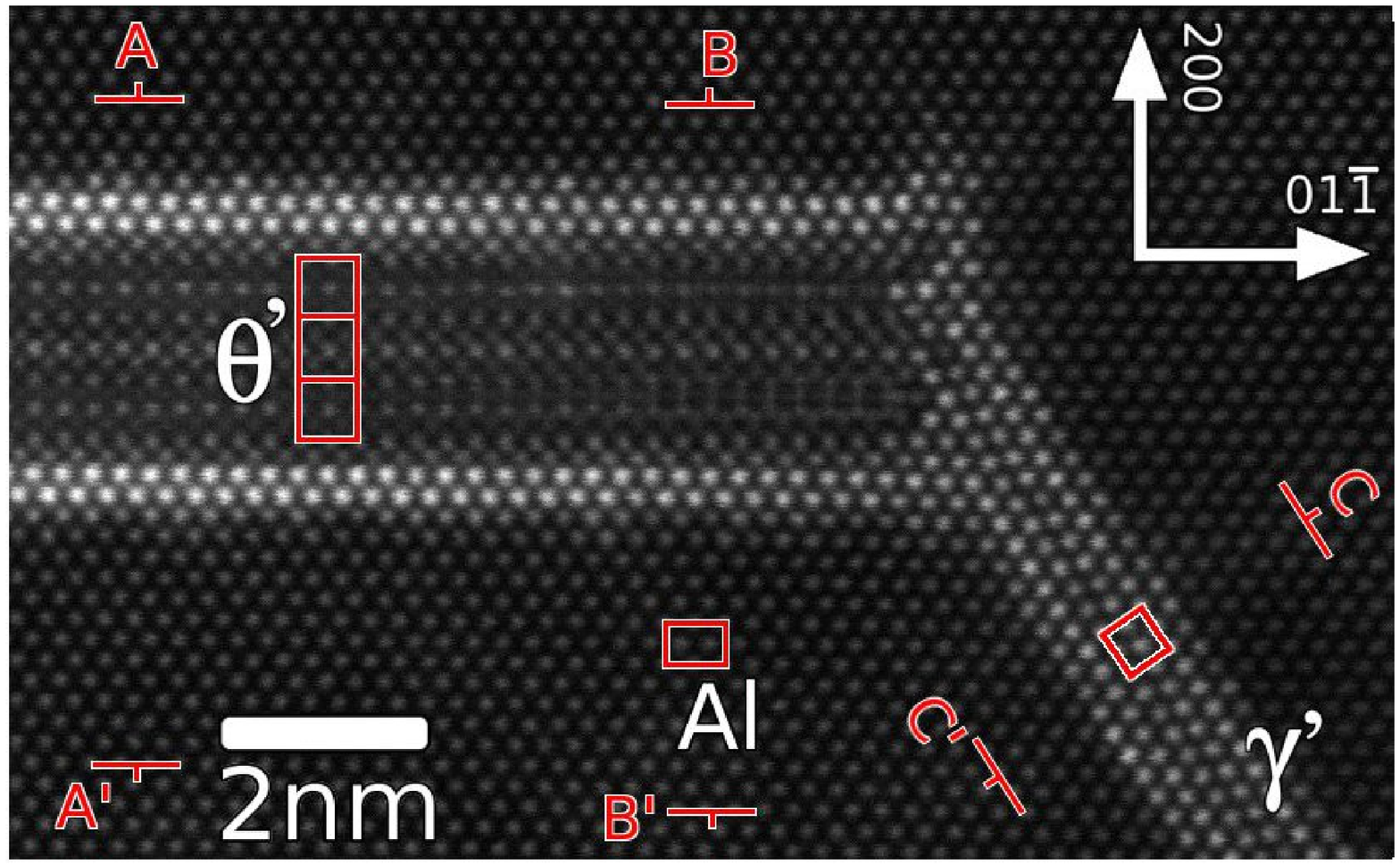}}
		\hfill
		\subfigure[HAADF-STEM\label{fig-theta-stem-2}]{\includegraphics[width=5.5cm]{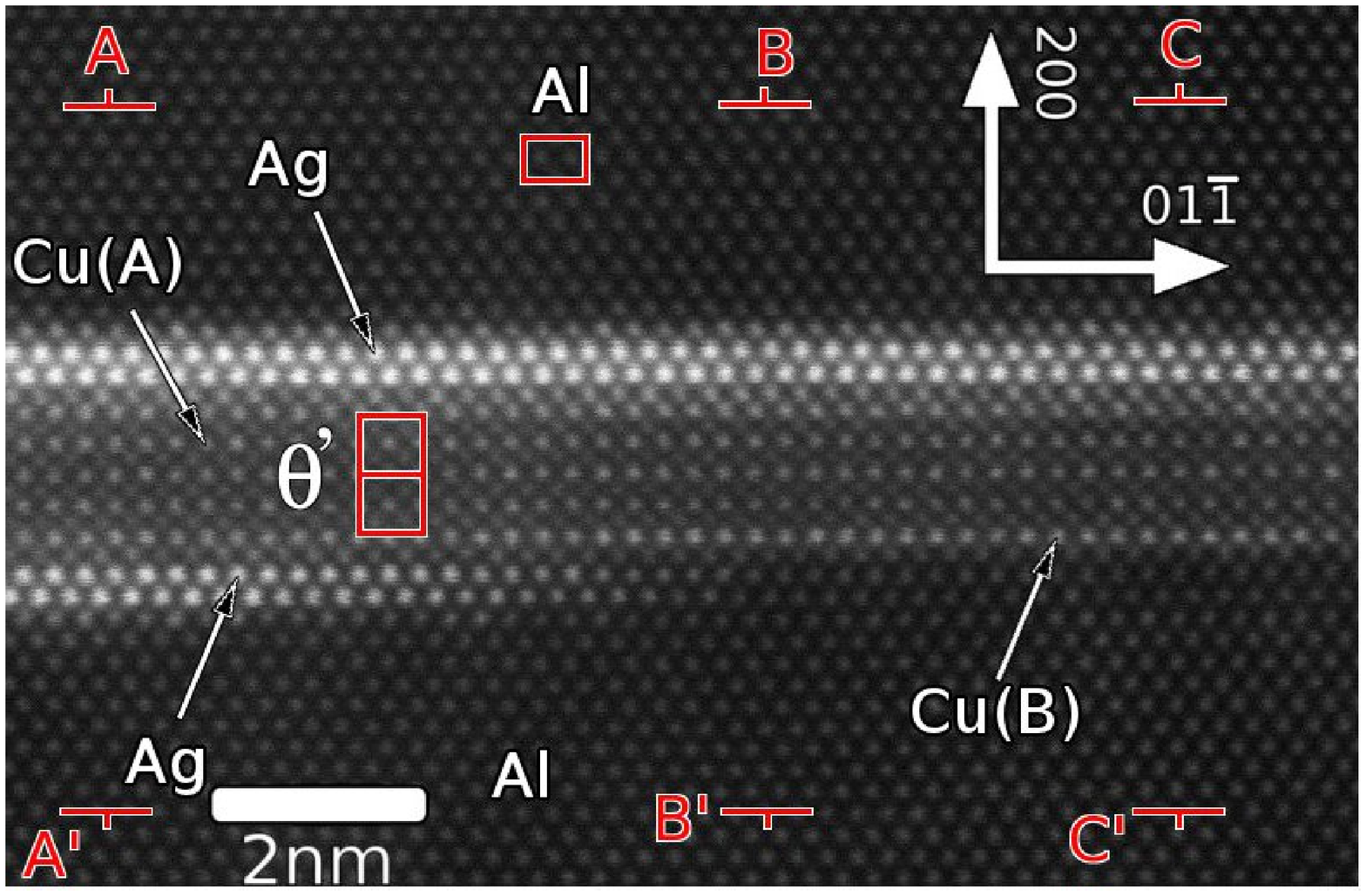}}
		\hfill\

		\hfill
		\subfigure[Intensity profiles from Fig. \ref{fig-theta-stem-1} \label{fig-theta-profile-1}]{\includegraphics[width=5.5cm]{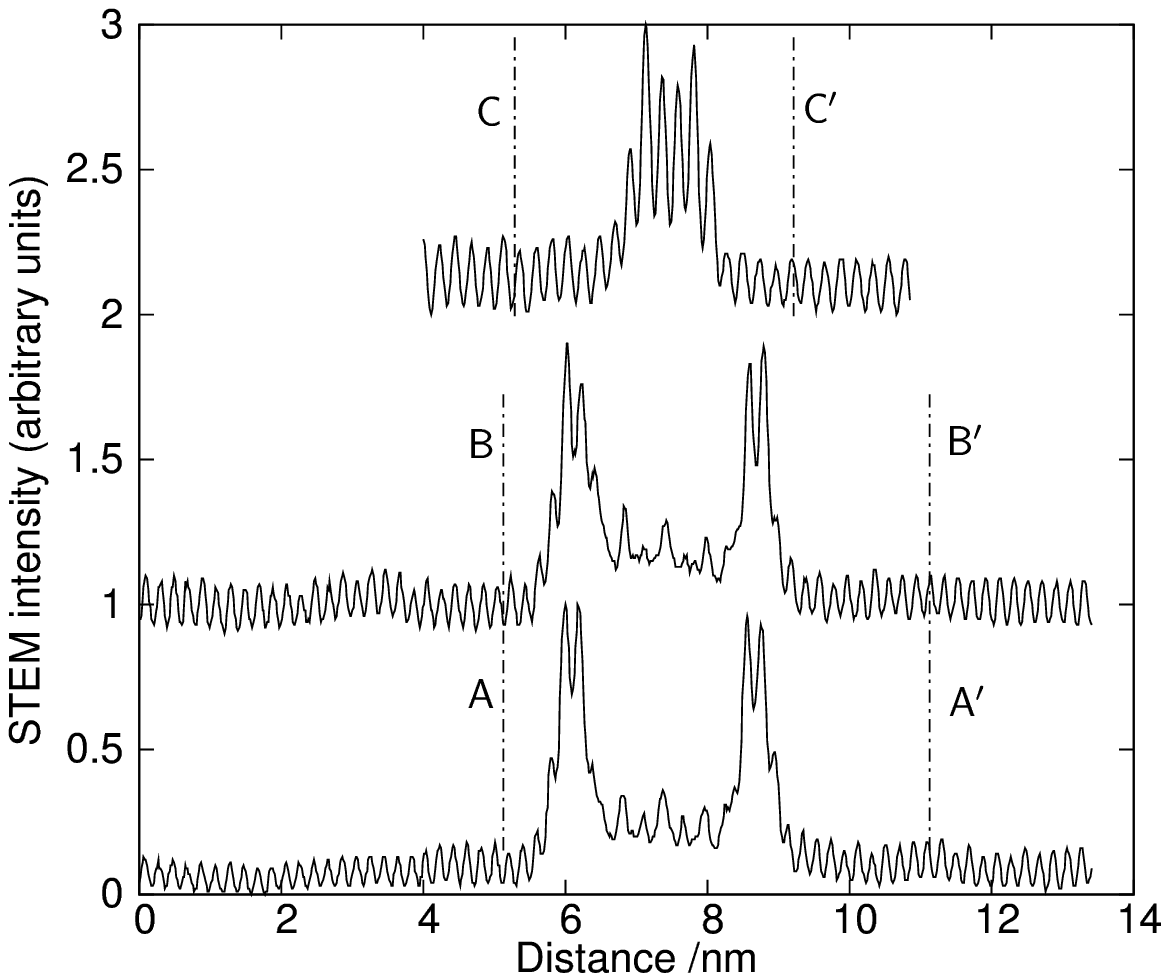}}
		\hfill
		\subfigure[Intensity profiles from Fig. \ref{fig-theta-stem-2}  \label{fig-theta-profile-2}]{\includegraphics[width=5.5cm]{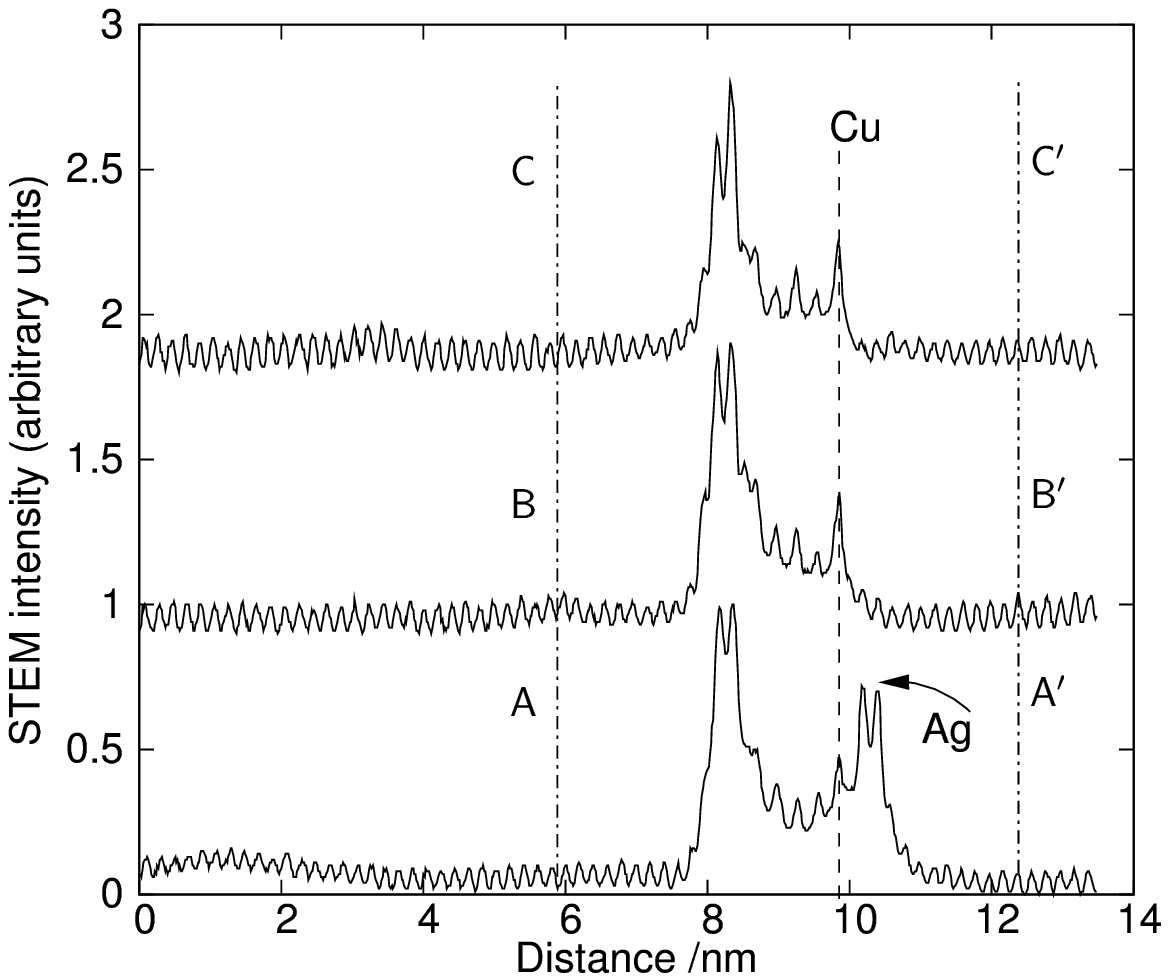}}
		\hfill\

		\caption{Ag enriched bi-layers on \thetaprime precipitates, viewed along the {[}011{]} zone axis.  
(a) and (b) show  HAADF-STEM images of \thetaprime precipitates with (a) complete and (b) partial Ag layers. 
There is a \gammaprime precipitate adjacent to the \thetaprime precipitate in (a).
(c) and (d) show intensity profiles of HAADF-STEM image intensity across A-A$^\prime$,B-B$^\prime$,C-C$^\prime$  in Figures (a) and (b), respectively.
 Al, \thetaprime and \gammaprime unit cells have been outlined near the label for each phase. 	 
\label{fig-theta-stem}}
	\end{center}
\end{figure*}

Isothermally ageing samples for 2-4\,h resulted in the precipitation of \thetaprime platelets at the dislocation loops. 
Figure~\ref{fig-theta-stem} shows representative HAADF-STEM micrographs of \thetaprime precipitates viewed along the  [011] zone axis.
Figure~\ref{fig-theta-stem-1} shows the coherent interfaces and semi-coherent edge of a \thetaprime platelet impinging on a \gammaprime precipitate. 
The coherent $(001)_{\theta^\prime}/(001)_\ce{Al}$ interfaces are surrounded by a bi-layer with strong atomic contrast. 
The HAADF-STEM intensity across the precipitate is shown in Fig~\ref{fig-theta-profile-1}. 
Intensity profiles have been drawn across the \thetaprime precipitate ($A$--$A'$ and $B$--$B'$) and  across the \gammaprime phase ($C$--$C'$). 
Peaks within the \thetaprime precipitate are due to  (002) Cu layers as the atomic columns in the intervening Al (002) planes are not well resolved. 
Profiles $A$ and $B$ show a strong double-peak just outside the \thetaprime phase (significantly greater in intensity than the Cu sites in \thetaprime) with the intensity decreasing to that of the matrix over 2-3 adjacent planes. 
The strong contrast in the interface layers relative to Cu sites in the \thetaprime phase (comparable in intensity to that of  the \gammaprime precipitate)  in combination with the Ag detected by EDX (Fig~\ref{edx-theta-Ag}) lead to the conclusion that these layers were dominated by silver (Z=47 \textit{vs.} Z=29 for Cu). 
The layer spacing up to and including the Ag-containing bi-layer does not diverge measurably from the (200) planar spacing in the matrix.

All \thetaprime precipitates observed in this alloy possess similar Ag-enriched coherent interfaces, although in some cases the Ag bi-layer was incomplete (e.g. Fig~\ref{fig-theta-stem-2}). 
The upper face of the \thetaprime precipitate has a uniform coverage of two strongly-contrasting atomic layers similar to that in Figure~\ref{fig-theta-stem-1}. 
On the bottom face this bi-layer gradually diminishes in intensity from right to left, with no clear variation in the spacing of the atomic columns.  
The HAADF image (Fig~\ref{fig-theta-stem-2}) and intensity profile (Fig~\ref{fig-theta-profile-2})  clearly show the reduction in intensity in the Ag rich layer from left ($A$--$A'$) to right ($C$--$C'$).

\begin{figure*}
	\begin{center}
		\hfill
		\subfigure[HAADF-STEM images of \thetaprime phase with Ag on both coherent interfaces. \label{fig-theta-stem-3}]{\includegraphics[width=5.5cm]{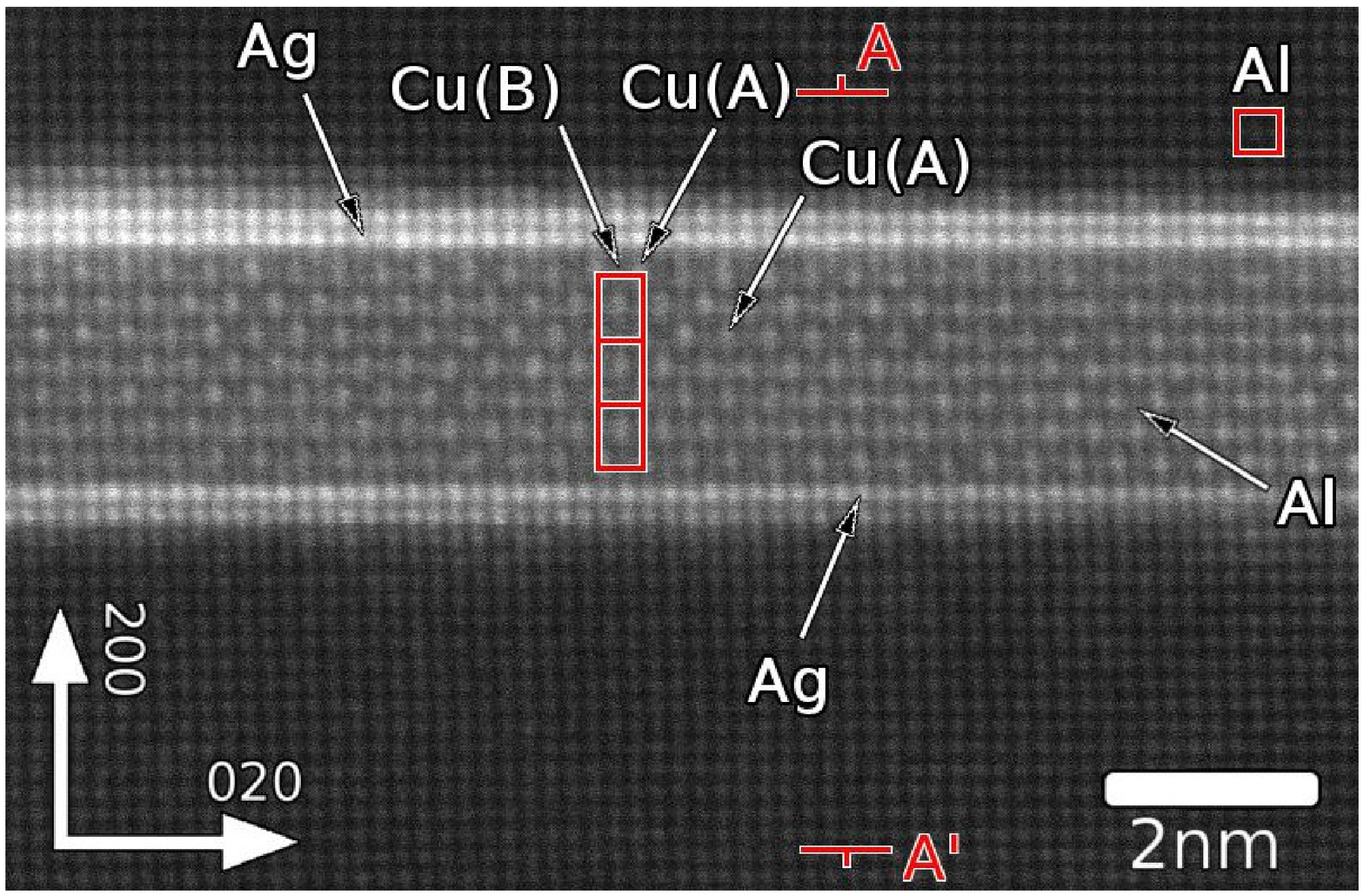}}
		\hfill
		\subfigure[HAADF-STEM image of \thetaprime phase with Ag on one interface.  \label{fig-theta-stem-4}]{\includegraphics[width=5.5cm]{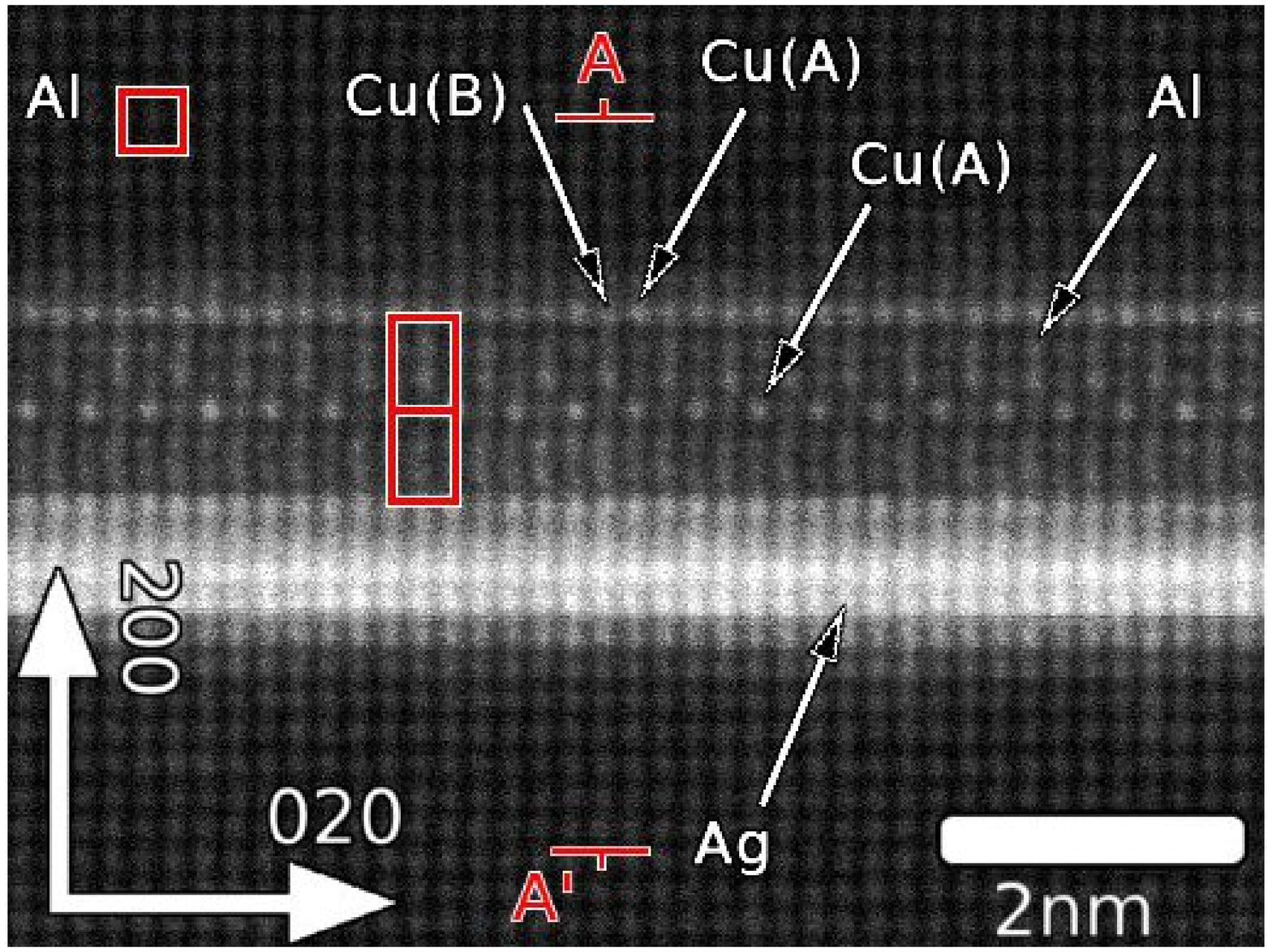}}
		\hfill\

		\hfill
		\subfigure[(Intensity profile from Fig~\ref{fig-theta-profile-3} \label{fig-theta-profile-3}]{\includegraphics[width=5.5cm]{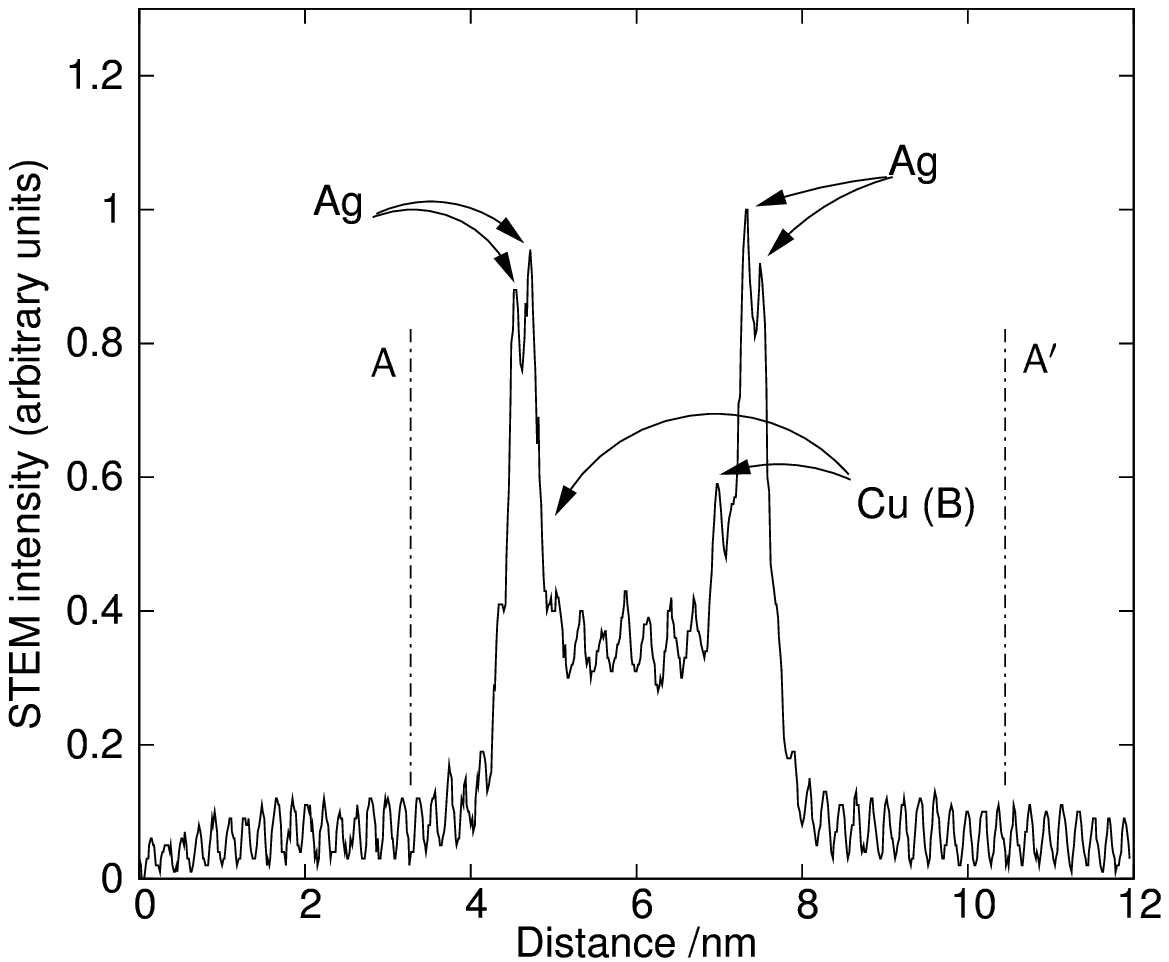}}
		\hfill
		\subfigure[Intensity profile from Fig~\ref{fig-theta-profile-4}\label{fig-theta-profile-4}]{\includegraphics[width=5.5cm]{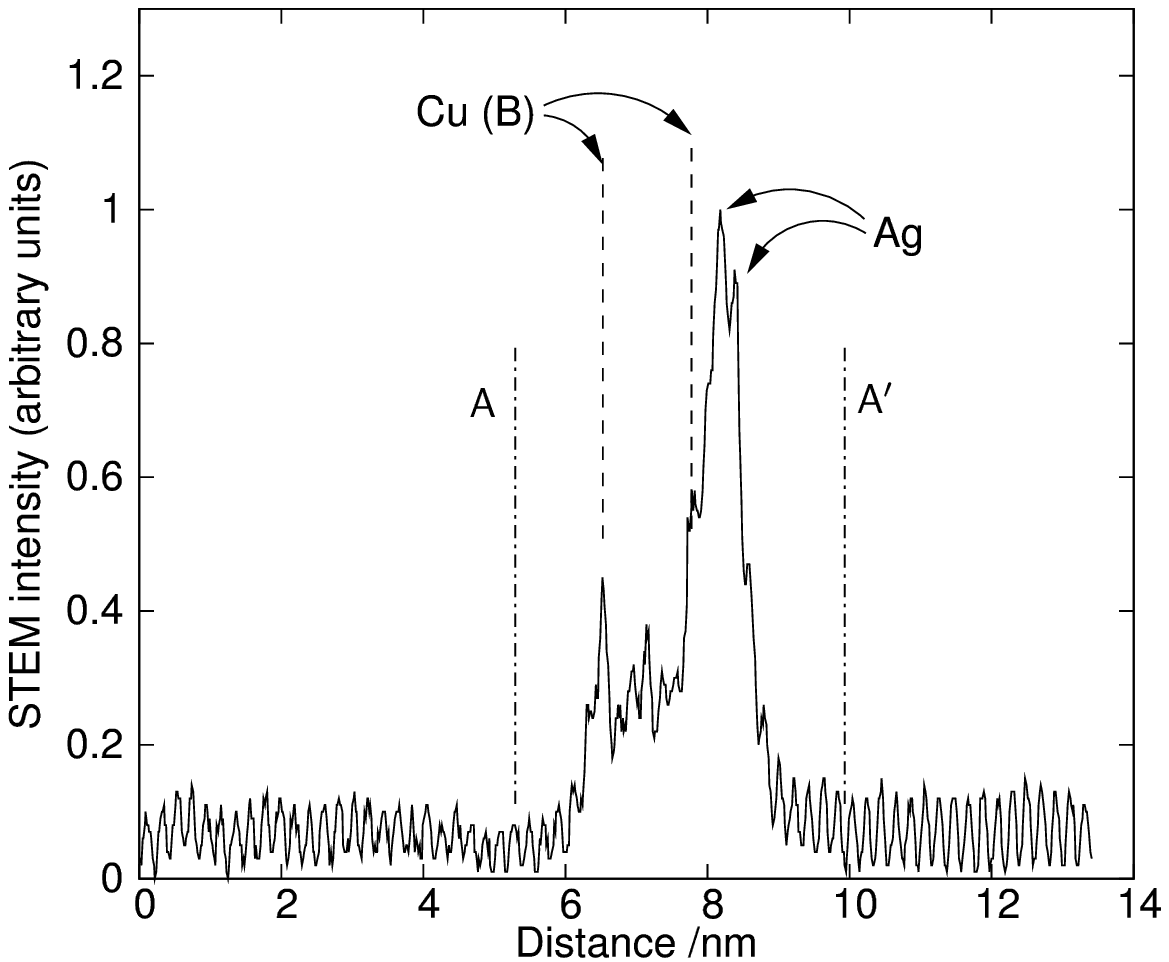}}
		\hfill\

		\caption{Ag enriched bi-layers on \thetaprime precipitates, viewed along the {[}001{]} zone axis.  
(a) and (B) show  HAADF-STEM images of \thetaprime precipitates with (a) near complete and (b) partial Ag bi-layers. 
There is a \gammaprime precipitate adjacent to the \thetaprime precipitate in (a).
(c) and (d) show  intensity profiles of HAADF-STEM intensity across the precipitates in Figures (a) and (b), respectively.
In (b) and (d) the stronger contrast in the outer layers of the \thetaprime structure (labeled Cu(B)) can be seen.
\thetaprime unit cells have been indicated by outlines. 
\label{fig-theta-profile}}
	\end{center}
\end{figure*}

The atomic positions of the Ag and Cu containing layers have been verified by imaging along a second direction. 
Figure~\ref{fig-theta-stem-3} shows a \thetaprime precipitate along the [001] zone axis, with strong contrast evident in two atomic layers on both faces of the precipitate, although the upper interface is brighter than the lower. 
Fig~\ref{fig-theta-stem-4} presents a example in which the Ag-rich bi-layer is absent on the upper interface and present on the lower interface, similar to Fig~\ref{fig-theta-stem-2}. 
In this image additional columns of Cu atoms can be seen on the outermost layer of the \thetaprime phase (labeled ``Cu(B)''),
which are of comparable intensity to columns in the bulk \thetaprime structure (``Cu(A)'').
HAADF-STEM intensity profiles drawn through the \thetaprime precipitates are shown in Fig~\ref{fig-theta-profile-3} and ~\ref{fig-theta-profile-4}.
In Fig~\ref{fig-theta-profile-3} the contrast of the lower matrix-\thetaprime interface is weaker than that of the upper interface.

The partial Ag layers in Fig~\ref{fig-theta-stem-2} and \ref{fig-theta-stem-4} allow  the Cu-enriched outer layers of the \thetaprime precipitate  to be seen clearly. 
When viewed along the  [011] direction, (Fig~\ref{fig-theta-stem-2})  the outermost Cu layers of the \thetaprime precipitate (labeled ``Cu(B)'') have stronger contrast than Cu sites in the bulk structure (``Cu(A)'').
This is evident in plots of the HAADF intensity across the precipitate (Fig~\ref{fig-theta-profile-2}). 
Where the Ag layer is absent  ($C$--$C^\prime$ in Fig~\ref{fig-theta-stem-2}) there is a single sharp peak corresponding to the Cu-enriched layer (labeled ``Cu(B)'').
The intensity of this peak is between that of the Ag enriched bi-layer and standard Cu planes of the precipitate and there is no indication of a shift in the peak position, regardless of whether Ag is present. 
When imaged along the [001] zone axis (Fig~\ref{fig-theta-stem-2})  the intensity of the interface sites is consistent with the bulk sites, but, additional atomic columns (``Cu(B)'') appear between the regular (``Cu(A)'') sites.  
Plots of HAADF-STEM intensity (Fig.~\ref{fig-theta-profile-4}) show a sharp single peak due to the additional Cu at the interface. 
These observations are consistent with a model for the introduction of additional Cu atoms at octahedral sites in \thetaprime at the Al-\thetaprime coherent interfaces \cite{BourgeoisAlCu2012}. 
Detailed examination reveals that similar additional Cu  atomic columns are present on the opposite, Ag-rich interfaces in Fig~\ref{fig-theta-stem-2} and \ref{fig-theta-stem-4} and are also present in Figures~\ref{fig-theta-stem-1} and \ref{fig-theta-stem-3}.

\begin{figure}
	\begin{center}
		\hfill
		\subfigure[]{\includegraphics[height=4.5cm]{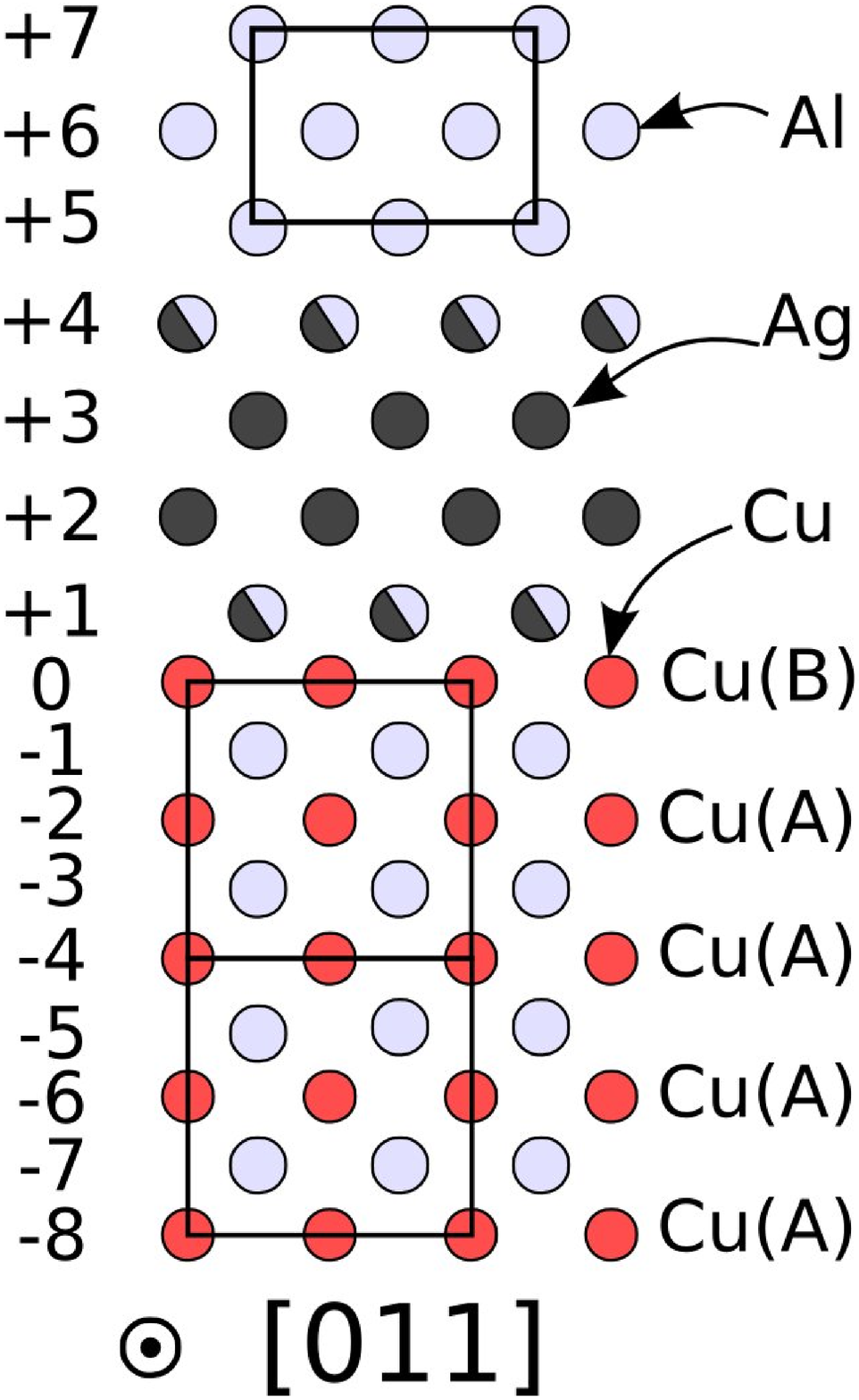}}
		\hfill
		\subfigure[]{\includegraphics[height=4.5cm]{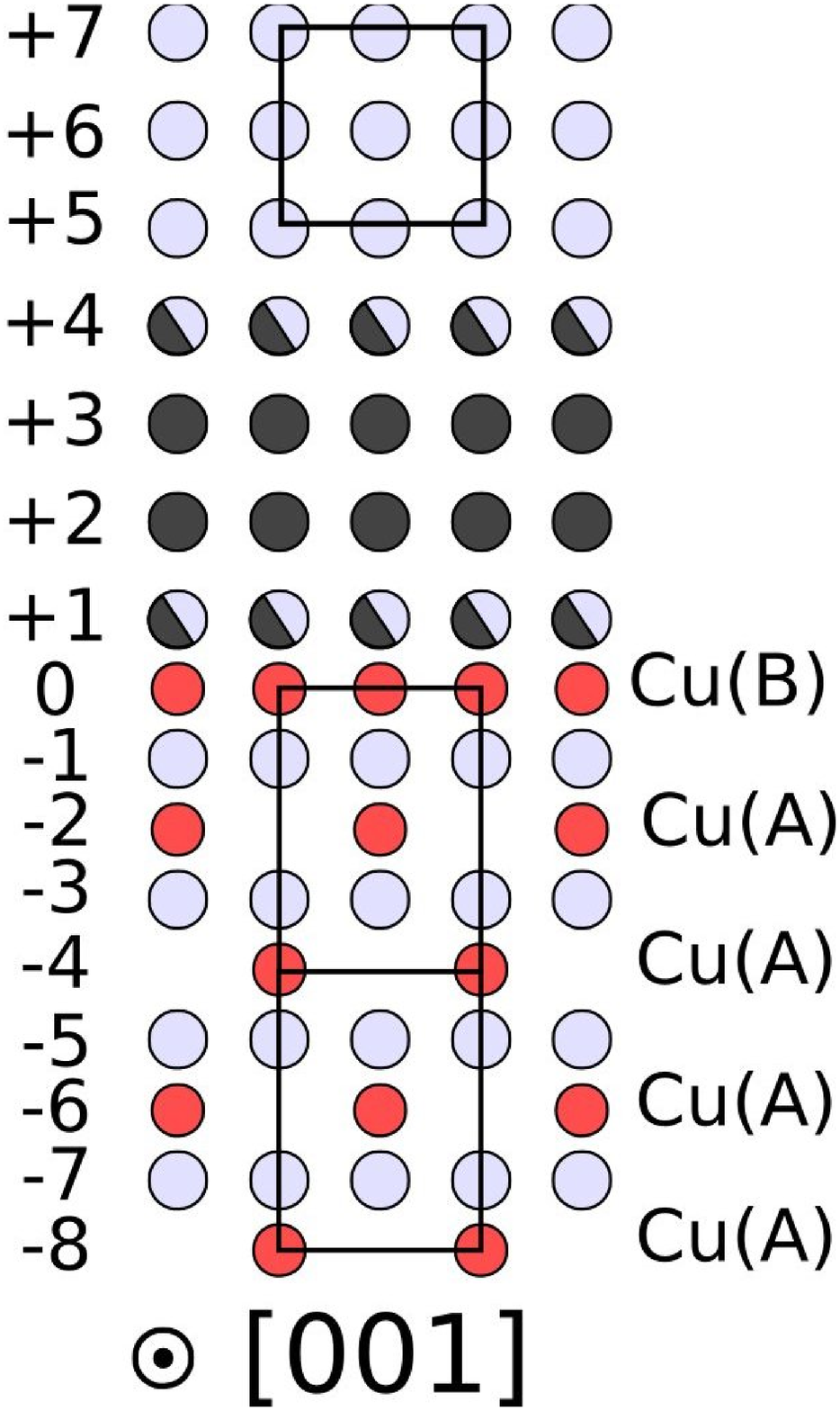}}
		\hfill\
		\caption{\label{fig-ledge-model}A structural model for the coherent \thetaprime-matrix interface in an Al-Cu-Ag alloy. 
Ag is concentrated into two layers, separated from \thetaprime by a single layer of mixed Al/Ag.
Cu layers with the bulk structure are designated Cu(A), while the interface layers with additional Cu in octahedral interstices are designated Cu(B). 
The solid outlines indicate Al and \thetaprime unit cells.}
	\end{center}
\end{figure}

A proposed structure for the coherent interface is illustrated in Figure~\ref{fig-ledge-model}.
Atomic layers  are numbered for convenience, commencing from the outermost Cu-rich layer of the \thetaprime precipitate. 
Cu layers labeled Cu(A) (Layers -2, -4, -6\dots) have the bulk structure.  
Those labeled Cu(B) (layer 0) have additional Cu in the octahedral interstices \cite{BourgeoisAlCu2012} which is responsible for stronger contrast along the [011] axis (Fig~\ref{fig-theta-stem}) and the appearance of additional atomic columns along the  [001] axis (Fig.~\ref{fig-theta-profile}).
Two layers outside the precipitate (labeled +2 and +3) are heavily enriched in silver, with the position of the atomic columns showing no measurable deviation from the position of the matrix atom columns along both [011] and [001] zone axis.
Although the location of these sites did not change, the HAADF-STEM intensity varies between \thetaprime precipitates  and between opposite faces and different regions of the same precipitate, suggesting a broad range of mixed Al-Ag occupancy.
Atomic sites in the adjacent layer (+1) are consistent with $fcc$ Al in the matrix and Al sites in the precipitate, as expected for a coherent \thetaprime-Al interface.
The atomic contrast in these columns, however, is much stronger, suggesting that this layer (and the opposite +4 layer)  were partially occupied by Ag, although at a lower concentration than the +2 and +3 layer.   
Cu occupancy of these sites is deemed unlikely given (a) the stronger atomic contrast, (b) the shorter Cu-Al bond length compared to Al-Ag or Al-Al and (c) the unfavourable Cu-Ag bond enthalpy, and (d) recently published EDX maps showing no detectable Cu surrounding the precipitate \cite{RosalieTms2012Thetaprime}.

\subsubsection{Semi-coherent \thetaprime-Al interfaces}

\begin{figure}
	\begin{center}
		\hfill
				\subfigure[{[}011{]} zone axis.   \label{fig-semi-coherent-1}]{\includegraphics[width=5cm]{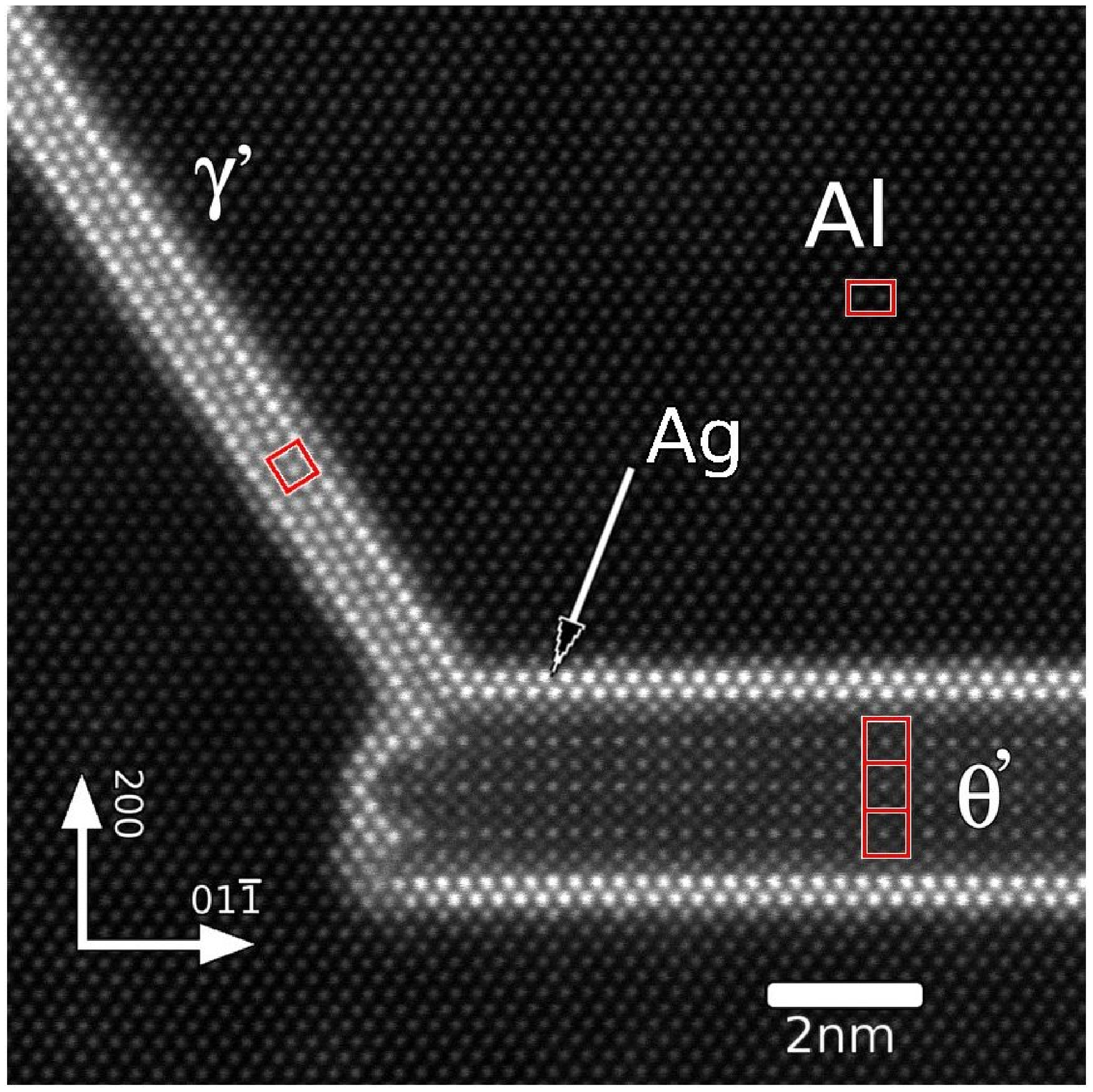}} \hfill\
				\subfigure[ {[}001{]} zone axis. \label{fig-semi-coherent-2}]{\includegraphics[width=5cm]{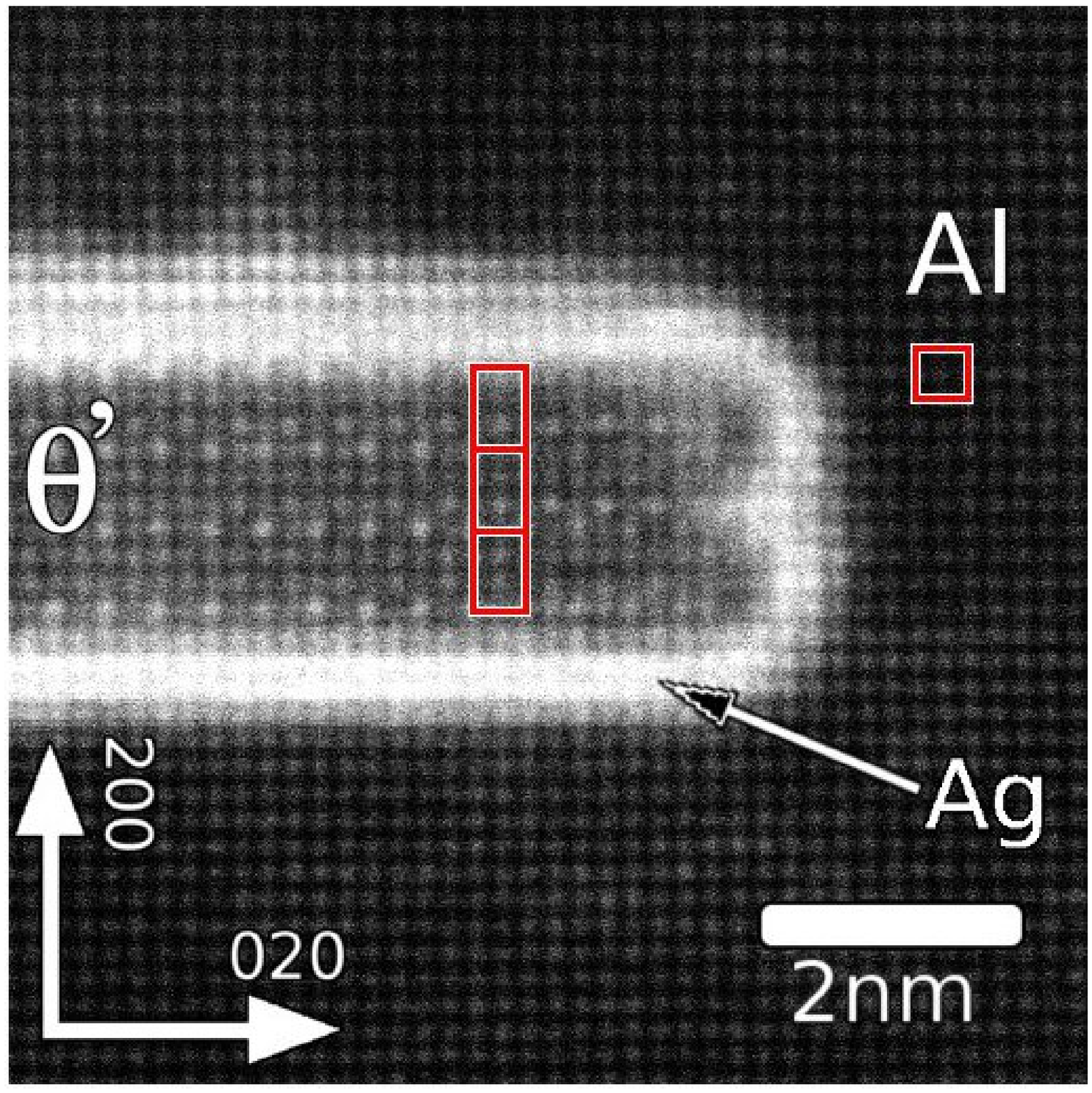}}		\hfill\
		\caption{Ag segregation to semi-coherent \thetaprime-Al interfaces. Foils were aged for 4h at 200$^\circ$C. 
Ag is concentrated in two atomic layers adjacent to the semi-coherent interface and in (a) shows facets parallel to \{111\} planes.
Al, \thetaprime and \gammaprime unit cells have been outlined. 
\label{fig-semi-coherent}}
 	\end{center}
\end{figure}

The semi-coherent \thetaprime-Al interfaces are frequently enriched in silver. 
Of a total of 40 instances in which the semi-coherent interface are clearly resolved, in 16 cases (40\%), the interfaces are completely covered by a double layer of silver.
In the remainder of the precipitates the Ag-rich bi-layer is either absent or incomplete. 
Figure~\ref{fig-semi-coherent} shows representative examples of semi-coherent edges of \thetaprime precipitates with Ag bi-layers along (a) [011] and (b) [001] zone axes.
In Fig.~\ref{fig-semi-coherent-1} the precipitate directly impinges on a \gammaprime precipitate. 
The semi-coherent interface is decorated with a bi-layer of Ag, similar to that present on the coherent interfaces.
The edge of the \thetaprime precipitate is jagged, with the adhering Ag bi-layer having facets that alternate between 
$(1\overline{1}1)$ and $(\overline{11}1)$ planes. 
It is notable that these are preferred habit planes for the \gammaprime (\ce{AlAg2}) phase, having the lowest interfacial energy \cite{Ramanujan1992a}.  
The semi-coherent interfaces are less well resolved along the [001] zone, (Fig.~\ref{fig-semi-coherent-2}); 
however, the Ag layer can still be seen to be two atomic layers in thickness.

\section{Discussion}
\subsection{\thetaprime phase precipitation in Al-Cu-Ag alloys}

Precipitation of the \thetaprime phase in Al-0.90at.\%Cu-0.90at.\%Ag alloys is observed on pre-existing assemblies of \gammaprime precipitates. 
Precipitation in the form of extended arrays or on dislocations is not observed  in this alloy, indicating either that the dislocation loops offer a more favourable nucleation site, or that the extended array morphology is prevented from forming. 
Recently reported diffraction contrast images of the \gammaprime precipitate assembly obtained before the nucleation of \thetaprime precipitates show diffuse, linear contrast features lying along $[1\overline{1}0]$ directions at either end of the assemblies \cite{RosalieTms2012Thetaprime}.
These have been shown to be segments of dislocation loops tangential to the (100) planes and dark-field imaging was used to show that the dislocation loop segment had dissociated  \cite{RosalieTms2012Thetaprime} as reported for precipitation of the \thetaprime phase colonies on dislocations \cite{dahmen:1983a} to yield a body-centred tetragonal stacking fault between  partial dislocations with Burgers vectors of type 1/2$\langle 001 \rangle$. 
Here it is shown using HAADF-STEM imaging and EDX mapping (Fig~\ref{edx-defect}) that Ag segregates to these defects, whereas the level of Cu was not measurably enhanced.  
Given that the dislocation is dissociated at this stage, despite the absence of Cu, it is evident that Ag segregation is responsible for the dislocation splitting. 
While there are no reports of Ag promoting dissociation of a dislocation in this way it has been established that the stacking fault energy  of Al close-packed (111) planes decreases locally for high levels of silver \cite{schulthess:1998}.  

\subsection{Silver segregation to \thetaprime-Al interfaces}

Interfacial energy, rather than strain, is the most probable cause for the formation of the Ag-rich bi-layers. 
The Al-Al, Ag-Ag bond lengths are 286.3\,pm and 288.9\,pm, respectively, and substitution of Ag for Al is not expected to relieve volumetric strain associated with the precipitate. 
In addition, measurable atomic displacements in atomic planes outside the \thetaprime structure are not seen in plots of the peak position (Fig~\ref{fig-theta-profile-2}, \ref{fig-theta-profile-4}), suggesting that the \thetaprime structure is not disrupted by the presence of the  Ag bi-layer. 

The generation of the Ag-rich bi-layer involves the loss of \thetaprime-Al and  \gammaprime-Al interfaces (as \gammaprime precipitates are removed) and the formation of a Al-Ag-Al-Cu interface.
Calculations of the interface energies for the Al-\thetaprime coherent precipitate-matrix interfaces have provided values\footnote{These calculations were based on the bulk structure, and did not include the addition Cu sites at the interface \cite{BourgeoisAlCu2012}. It is unlikely, however that this would substantially affect the interfacial energy.}
ranging from 156\,mJ.m$^{-2}$ using molecular dynamics \cite{HuInterface2006} to 190\,mJ.m$^{-2}$  for some first-principles calculations \cite{Vaithyanathan2004}.
Discrete lattice plane method calculations for  coherent \gammaprime-Al \{111\} interfaces estimated a much lower energy of  6.5\,mJ.m$^{-2}$, with facets on other planes in the range  6.3-7.1\,mJ.m$^{-2}$ \cite{Ramanujan1992a}.
Wetting of the \{001\} \thetaprime-Al coherent interface would therefore reduce the interfacial energy with the matrix by replacing the higher energy \thetaprime-Al interface with a Ag-Al interface.

\begin{table}
	\begin{center}
	\caption{Bond enthalpies (kJ/mol) for gaseous diatomic species \cite{CrcHandbook2005}. \label{tab-enthalpy}}
	\begin{tabular}{llll}
	\toprule
	& Al & Ag & Cu \\ \midrule
Al	& 133$\pm$5.8		& 183.7$\pm$9.2	&	227.1$\pm$1.2 		\\
Ag	& 				&159$\pm$2.9		&	171.5$\pm$9.6	\\
Cu	& 				& 				&	201\\
\bottomrule
	\end{tabular}
	\end{center}
\end{table}

The existence of a single atomic layer between a precipitate and the surrounding Ag interface layers was also observed for $\Omega$ precipitates in Al-Cu-Mg-Ag alloys \cite{HutchinsonOmega2001}.
According to density functional theory calculations \cite{SunOmega2009}, Mg occupancy of this layer gave the lowest energy structure; however, it was reported that there was no clear indication of additional Mg in or around the precipitates \cite{HutchinsonOmega2001}, which would infer that this layer contains Al.
In the present alloy, Mg is a trace-level impurity and this layer undoubtedly consists principally of Al, with the stronger atomic contrast (See Fig~\ref{fig-theta-stem},\ref{fig-theta-profile}) suggesting partial Ag occupancy.

The presence of Al atoms in this layer is probably due to the relative bond enthalpies\footnote{Bond enthalpies are given as the values for gaseous diatomic species to provide a qualitative comparison of the relative bonding strength.}.
Although atomistic calculations such as by density functional theory are required to confirm this view, the following explanation appears plausible. 
In the absence of this layer, the Cu-terminated \thetaprime interface would be in contact with Ag 
(bond enthalpy for (171.5$\pm$9.6\,kJmol$^{-1}$).
With this layer in place the system is able to form stronger Al-Cu (227.1$\pm$1.2\,kJmol$^{-1}$) and Al-Ag 
183.7$\pm$9.2\,kJmol$^{-1}$) bonds (see Table~\ref{tab-enthalpy}).

There appears to be no structural or chemical alteration in the \thetaprime phase, despite the presence of the Ag bi-layer.
The outermost layer of the \thetaprime phase even contains additional Cu atoms in octahedral sites, as has recently been demonstrated in binary Al-Cu \cite{BourgeoisAlCu2012}. 
This points to an interaction in which Ag chemically wets the Al-\thetaprime interface, given sufficient solute supply, lowering the interfacial energy, without substantially affecting the \thetaprime phase. 

A substantial amount of Ag is required to form a interface bi-layer and larger  \thetaprime precipitates with partial Ag coverage are observed (e.g. Figs~\ref{fig-theta-stem-3} and \ref{fig-theta-stem-4}), suggesting that the Ag supply can become deficient. 
In an earlier study on Al-Cu-Ag aged for 16$-$72\,h \cite{rosalie:2005} the \gammaprime assemblies were rarely seen. 
The most prominent feature were paired  \thetaprime precipitates,  separated by roughly the same spacing as the width of the \gammaprime assemblies with few remaining \gammaprime precipitates.
It therefore appears likely that silver is drawn not only from solution and GP zones, but from the neighbouring \gammaprime precipitates, leading to the disruption and eventually consumption of the precipitate assembly. 
This in itself suggests a strong chemical interaction for the interface Ag to be preferred over the \gammaprime structure. 

Some 40\% of semi-coherent \thetaprime interfaces are completely covered with Ag.  
The Ag bi-layer followed the shape of the precipitate edge closely, with a tendency for faceting along \{111\} planes. 
Discrete lattice plane calculations show that the $\{111\}_{\mathrm{Al}}/ (0001)_{\gammaprime}$  interface had an extremely low chemical interfacial energy of 5-7mJ.m$^{-2}$ \cite{Ramanujan1992a}. 
The presence of such a silver bi-layer seems to preclude further lengthening of the \thetaprime precipitate, since Cu solute would have to diffuse through 2 atomic layers of silver, with which it has a weaker bond enthalpy than with Al (227.1 \textit{vs.} 171.5\,kJmol$^{-1}$).

It is unclear whether the Ag-enriched bi-layer affects the preferred thicknesses \cite{StobbsPurdy1978} or thickening behaviour of the \thetaprime phase.
An investigation into the thickening behaviour of \thetaprime precipitates in Al-Cu-Ag is underway. 

\subsection{Comparison with other Al alloy systems}
This study highlights that the details of the precipitate-matrix interface are strongly system dependent and that the matrix or  precipitate may undergo compositional and structural changes. 
In Al-Cu binary alloys, it is the structure of the \thetaprime precipitate that differs from that of the bulk phase, with additional Cu atoms present at the interface  \cite{BourgeoisAlCu2012}.  
Silicon additions to Al-Cu alloys also affect the precipitate composition. 
Atom-probe measurements found a factor of $>$11 increase in the Si content at the coherent \thetaprime-Al interface in Al-4at.\%Cu-0.02at.\%Si  alloys, which density functional theory calculations attributed to Si substituting for Cu in the precipitate \cite{BiswasThetaprime2011}.

In alloys containing $\Omega$ phase precipitates,  it is the matrix composition that changes, with the formation of a Ag/Mg interface bi-layer \cite{OkunishiOmega2001,HonoOmega1993,HutchinsonOmega2001}.
The Ag-enriched bi-layers shown here in Al-Cu-Ag alloys bear striking similarities to the Mg/Ag bi-layers surrounding $\Omega$ phase  precipitates. 
However, there was no evidence of an interface layer on the semi-coherent edges of $\Omega$ precipitates (see, for example Fig.~5 and 8 in \cite{HutchinsonOmega2001}),  whereas such layers is frequently observed on \thetaprime precipitates (e.g. Figs~\ref{fig-theta-stem-1} and \ref{fig-semi-coherent}). 
It should also be noted that density functional theory calculations predict that only a combination of Mg and Ag stabilises the  $\Omega$-Al interface and that charge transfer from Mg may play a role in this effect \cite{SunOmega2009}. 
In this study,  Mg was present only as an impurity element at levels of $<$0.005wt.\% ($\approx$0.005at.\%, compared to 0.90at.\%Ag) and it appears very unlikely that Mg plays any role in the bi-layers in Al-Cu-Ag alloys. 

The observation of additional Cu in the outermost layer of the \thetaprime phase shows that this phenomenon is not unique to \thetaprime precipitates in binary Al-Cu and can exist despite the presence of a Ag interface bi-layer. 
In Al-Cu-Ag alloys, therefore, both the matrix and precipitate undergo changes at the interface. 
This highlights the complexity of interface interactions and the possibility for compositional or structural deviations in the matrix (Ag bi-layers in Al-Cu-Ag, Ag/Mg bi-layers in  Al-Cu-Mg-Ag) and/or precipitate (extra Cu in \thetaprime in Al-Cu and Al-Cu-Ag, $\Omega$ precipitation in Al-Cu-Mg-Ag, Si substitution in \thetaprime Al-Cu-Si).

Finally, it should be noted that the appearance of the thin \gammaprime precipitates was also of interest. 
Single unit cell thickness \gammaprime precipitates   (Fig.~\ref{defect-HAADF-STEM}, \ref{fig-theta-profile-1} and \ref{fig-semi-coherent-1}) did not show the alternating strong and weak contrast that was expected for the Ag-rich and Al-rich layers in this structure \cite{howe:1987a,zarkevich:2002}. 
In an earlier report on thickening of \gammaprime precipitates, we noted that Ag and Al did not appear to be ordered in single unit cell thickness \gammaprime precipitates \cite{RosalieActa2011} and an investigation into ordering of silver in  \gammaprime precipitates in Al-Cu-Ag alloys is underway. 

\section{Conclusions}

\thetaprime (\ce{Al2Cu}) precipitates in Al-Cu-Ag alloys were examined using EDX mapping and high resolution HAADF-STEM imaging. 
Silver (Ag) played several roles in the nucleation and growth of \thetaprime precipitates on dissociated segments of dislocation loops.

\begin{enumerate}
\item Ag in solid solution facilitates dissociation of the dislocation loop to provide an isostructural surface between 
$1/2 \langle 100\rangle$-type partial dislocations, assisting in the nucleation of the \thetaprime phase. 

\item Ag segregates to the coherent interfaces of the \thetaprime precipitates, forming an Ag-rich bi-layer.
This bi-layer is thought to reduce the chemical component of the interfacial energy between \thetaprime and the matrix. 
As \thetaprime precipitates lengthen, Ag  is drawn from the adjacent regions and neighbouring \gammaprime 
(\ce{AlAg2}) precipitates,  leading to the gradual shrinkage and loss of the \gammaprime precipitates.
\item Ag also segregated to approximately 40\% of the semi-coherent \thetaprime-matrix interfaces, where it appears to act as a barrier to Cu diffusion,  thus impeding or preventing lateral growth of the \thetaprime precipitates. 
\end{enumerate}

\section*{Acknowledgments}
The authors gratefully acknowledge the support of the Australian Research Council through the Centre of Excellence for Design in Light Metals.
The experiments were conducted at the Monash Centre for Electron Microscopy and we acknowledge the use of the facilities and engineering support by Russell King.
We also thank Dr. Brian Pauw for valuable comments on the manuscript. 
Finally we are thankful for the support and encouragement of Professor Barrington C. Muddle, whose work initiated this project.

\end{document}